\documentclass[twocolumn,aps,prb,citeautoscript,amsmath,amssymb,floatfix,eqsecnum]{revtex4-2}
\usepackage{amsmath,amssymb,color}
\usepackage{bm}
\usepackage{graphicx}
\usepackage{psfrag}
\usepackage{hyperref}

\begin{document}

\title{Non-Fermi liquid fixed point 
of the  dissipative  
Yukawa-Sachdev-Ye-Kitaev model
}

\author{Niklas Cichutek, Andreas R\"{u}ckriegel, Max O. Hansen, and Peter Kopietz}
  
\affiliation{Institut f\"{u}r Theoretische Physik, Universit\"{a}t
  Frankfurt,  Max-von-Laue Stra{\ss}e 1, 60438 Frankfurt, Germany}

\date{April 1, 2024}

 \begin{abstract}

Using a functional renormalization group approach we derive the renormalization group (RG) 
flow of a dissipative variant 
of the Yukawa-Sachdev-Ye-Kitaev model  
describing $N$ fermions on a quantum dot which interact via a disorder-induced Yukawa
coupling with $M$ bosons. The inverse Euclidean propagator of the bosons is assumed to exhibit 
a non-analytic term  
proportional to the modulus
of the Matsubara frequency. 
We show that, to leading order in $1/N$ and $1/M$, 
the hierarchy of formally exact
flow equations for the irreducible vertices 
of the disorder-averaged model 
can be closed at the level of
the two-point vertices.
%
We find that the RG flow exhibits a non-Fermi liquid fixed point characterized by a finite fermionic anomalous dimension $\eta$ which is related to the bosonic anomalous dimension $\gamma$ via the scaling law $2 = 2 \eta + \gamma$ with $ 0 < \eta < 1/2$.
We explicitly calculate $\eta$ and the critical exponents characterizing the linearized
RG flow in the vicinity of the fixed point as functions of $N/M$.
\end{abstract}


\maketitle


\section{Introduction}

The Yukawa-Sachdev-Ye-Kitaev (YSYK) model describes $N$ fermions on a quantum dot 
which are coupled to $M$ phonons via infinite-range random 
couplings~\cite{Esterlis19,Hauck20,Wang20a,Wang20b,Pan21,Classen21,Davies22,Valentinis23a,Valentinis23b}.
Recently this model has also attracted attention because it features maximally chaotic behavior~\cite{Davies22,Kim20}.
In the limit $N \rightarrow \infty$ and
$M \rightarrow \infty$ with $N/M = {\cal{O}} (1)$ 
(which we call for simplicity large-$N$ limit) 
the YSYK model exhibits a non-Fermi liquid state 
which can be studied in a controlled way. 
In this limit
the perturbative expansion of the fermionic and bosonic self-energies 
is dominated by so-called melon-diagrams 
which can be summed to all orders in perturbation theory 
by solving coupled Dyson-Schwinger (DS) equations. 
Alternatively, 
the DS equations can be derived from the large-$N$ saddle point 
of a suitably defined functional integral involving
special bi-local composite fields \cite{Sachdev15,Kitaev17,Esterlis19}.
From the numerical solution of the DS equations for the YSYK model Pan {\it{et al.}}~\cite{Pan21} have shown that the model exhibits
for vanishing chemical potential $\mu$
a non-Fermi liquid phase where the fermionic and bosonic 
self-energies display power-law behavior 
characterized by exponents which depend on the ratio $N/M$. Moreover, 
the power-law behavior of the self-energies  
persists even for finite values of $\mu$, a phenomenon which has been called {\it{self-tuned}} 
criticality \cite{Pan21}.
The leading power-law behavior in the non-Fermi liquid phase 
has also been extracted analytically by self-consistently solving
the DS equation at low energies \cite{Classen21}. 
However, the physical reason 
for the self-tuned criticality
of the non-Fermi liquid phase of the YSYK model
has not been identified. 

From  the point of view of the 
renormalization group (RG) the phenomenon of 
self-tuned criticality has a simple explanation:  
the critical state must be associated with a 
fixed point with only
attractive directions. Such a fixed point is called a 
sink \cite{Goldenfeld92,Kopietz10,Smit21} and describes a stable phase of matter.
In fact, the
non-Fermi liquid phase of the Sachdev-Ye-Kitaev model with complex fermions has recently been shown to be associated with such a sink~\cite{Smit21}. 
Unfortunately, we have not been able to obtain sensible results for the 
critical behavior of the YSYK model using a straightforward generalization 
of the functional renormalization group (FRG) approach developed 
in our previous work \cite{Smit21} for the SYK model.
The reason for this failure of the standard FRG approach are not entirely 
clear to us at this point; possible explanations are the failure of the low-energy expansion or  subtleties associated with the proper regulator choice in
coupled Fermi-Bose systems with two types of frequency 
scaling~\cite{Schuetz05}.
It turns out, however, that these technical complications
do not arise for the dissipative variant of the YSYK model
where the frequency-dependence of the inverse boson propagator
is proportional to the modulus $| \Omega |$ of the 
bosonic Matsubara frequency. 
In this work we will therefore focus 
on this dissipative YSYK model and use a generalization of the
FRG approach developed in Ref.~[\onlinecite{Smit21}] to investigate the
RG flow of this model. 
Our motivation for studying this model is not solely technical, 
because the dissipative YSYK model can be viewed as a toy model 
for understanding the behavior of strongly correlated
fermions that are coupled to bosons with dissipative dynamics.

We define the dissipative YSYK model via the following
Euclidean action,
 \begin{equation}
 S = S_2 + \frac{1}{\sqrt{MN}}
\sum_{ijk} g_{ i j k} \sum_{\sigma}  \int_0^{\beta} d \tau   
\bar{c}_{i \sigma} ( \tau )  {c}_{j \sigma} (\tau ) \phi_k ( \tau )  ,
 \label{eq:Sdef}
 \end{equation}
where $\beta$ is the inverse temperature, $\phi_k ( \tau )$ is a real bosonic field depending on a flavor index $k$ and
imaginary time $\tau$,
$c_{i \sigma} ( \tau )$ and $\bar{c}_{i \sigma} ( \tau )$ are Grassmann variables
representing electrons of type $i$ with spin projection $\sigma$, and
the quadratic part of the action is in frequency space given by
 \begin{align}
 S_2 = {} &   -   \frac{1}{\beta} \sum_{\omega}    \sum_{i=1}^N \sum_{\sigma}
 ( i \omega +  \mu ) \bar{c}_{ i \sigma \omega} c_{i \sigma \omega}
 \nonumber
 \\
 & + \frac{1}{2 \beta} \sum_{\Omega} \sum_{k=1}^M   (| \Omega |  + \Delta ) \phi^{\ast}_{ k \Omega} \phi_{ k \Omega}.
 \label{eq:S2def}
 \end{align}
Here $\Delta$ defines the bare energy scale of the bosons,
$\omega$ and $\Omega$ are fermionic and bosonic Matsubara frequencies, and the
Fourier components of the fields are defined by
 \begin{equation}
 c_{ i \sigma \omega} = \int_0^{\beta} d \tau e^{  i \omega \tau } c_{ i \sigma} ( \tau ),
 \; \; \;
 \phi_{ k \Omega} =  \int_0^{\beta} d \tau e^{  i \Omega \tau } \phi_k ( \tau ).
 \end{equation}
The Yukawa couplings
$g_{ij k}$ in Eq.~\eqref{eq:Sdef}
are independent Gaussian random variables with vanishing average and constant
variance. In general the couplings
$g_{ij k} = g^{\prime}_{ ijk} + i g^{\prime \prime}_{ ijk}$ are complex.
The hermiticity of the Hamiltonian implies that the real and imaginary parts have the
symmetries $ g^{\prime}_{ij k} = g^{\prime}_{ ji k}$ and
 $ g^{\prime \prime}_{ij k} = - g^{\prime \prime}_{ ji k}$.
A finite imaginary part $g^{\prime \prime}_{ijk}$ is a manifestation of 
broken time-reversal symmetry \cite{Classen21}. 
The second moments of the couplings are assumed to be of the 
form \cite{Classen21}
 \begin{subequations}
 \label{eq:secondmom}
 \begin{align}
 \langle g^{\prime}_{ i j k} g^{\prime}_{ i^{\prime} j^{\prime} k^{\prime} } \rangle
 & = g_1^{ 2} ( \delta_{ i i^{\prime}} \delta_{ j j^{\prime}}
 + \delta_{ i j^{\prime} } \delta_{ j i^{\prime} } ) \delta_{ k k^{\prime} },
 \\
 \langle g^{\prime \prime}_{ i j k} g^{\prime \prime}_{ i^{\prime} j^{\prime} k^{\prime} } \rangle
 & = g_2^{ 2} ( \delta_{ i i^{\prime}} \delta_{ j j^{\prime}}
 -  \delta_{ i j^{\prime} } \delta_{ j i^{\prime} } ) \delta_{ k k^{\prime} },
 \\
 \langle g^{\prime}_{ i j k} g^{\prime \prime}_{ i^{\prime \prime} j^{\prime} k^{\prime} } \rangle
 & = 0,
 \end{align}
 \end{subequations}
where $\langle \ldots \rangle$ denotes averaging over the Gaussian 
probability distribution of the random couplings.
The non-analytic  $| \Omega |$-dependence in the bosonic part of $S_2$
describes dissipative bosonic dynamics 
due to the coupling to some other degrees of freedom that do not 
explicitly appear in the above action. Note that the 
usual YSYK model can be obtained by replacing $| \Omega | \rightarrow \Omega^2$
in Eq.~\eqref{eq:S2def}.

The rest of this work is organized as follows. 
In Sec.~\ref{sec:FRG} 
we derive FRG flow equations for the irreducible vertices of the disorder-averaged
YSYK model and develop a truncation of the hierarchy of FRG flow equations 
which becomes exact for $N \rightarrow \infty$ and $M \rightarrow \infty$.
The resulting flow equations are not precisely  equivalent to the 
DS equations~\cite{Esterlis19,Hauck20,Wang20a,Wang20b,Pan21,Classen21,Davies22,Valentinis23a,Valentinis23b} for the YSYK model. However, the DS equations can be recovered if we modify our
flow equations using the so-called Katanin substitution \cite{Katanin04}.
In Sec.~\ref{sec:flow} we simplify our FRG flow equations for the irreducible self-energies
using  a standard low-energy expansion which reduces our functional flow equations to a system of
ordinary differential equations for five scale-dependent couplings associated with the
chemical potential, the boson gap, and three types of wave-function renormalization factors.
In Sec.~\ref{sec:fixedpoint} 
we show that our flow equations have a non-trivial 
non-Fermi liquid fixed point. 
We derive the linearized RG flow in the vicinity of this fixed point
and show that it has only one repulsive direction corresponding to
a linear combination of the rescaled chemical potential and a parameter describing
the spectral asymmetry. All other couplings are irrelevant at this fixed point, so that fine-tuning
of the bosonic energy scale $\Delta$ in Eq.~\eqref{eq:S2def} is not necessary to realize the
non-Fermi liquid phase. 
Finally, in
Sec.~\ref{sec:conclusions} we present our conclusions.

\section{Large-$N$ truncation of the exact RG flow equations}
 \label{sec:FRG}

\begin{figure}[tb]
 \begin{center}
  \centering
\vspace{7mm}
  \includegraphics[width=0.45\textwidth]{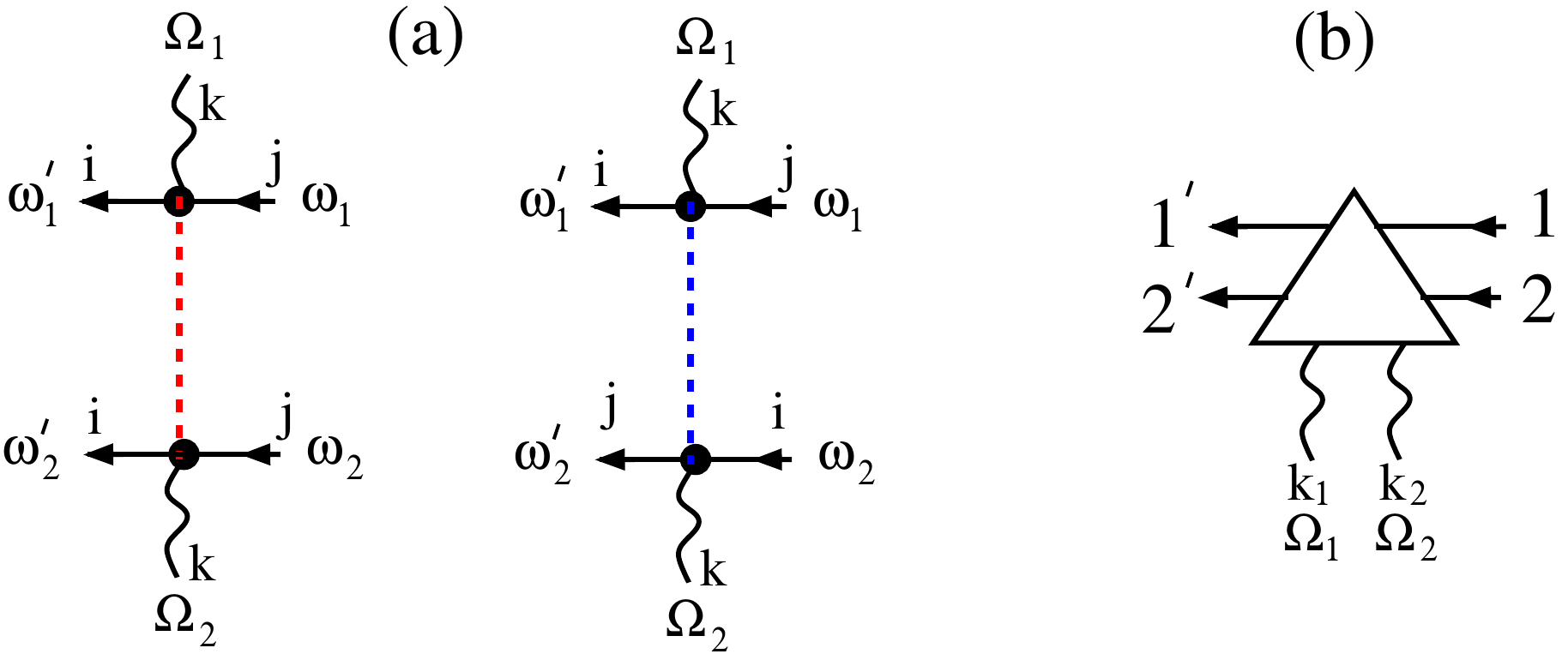}
   \end{center}
  \caption{
(a) Graphical representation of the bare interaction vertices in the bare action 
$S_6$
given in Eq.~\eqref{eq:S6freq}.  Outgoing arrows represent $\bar{c}_{ i \sigma \omega}$, 
incoming arrows represent $c_{i \sigma \omega}$,  
wavy external lines
represent $ \phi_{ k \Omega} $,
the red dashed line represents the bare 
interaction vertex
$ g_1^2 - g_2^2$ while the blue dashed line represents $g_1^2 + g_2^2$. 
Note that on both ends of the dashed lines
the frequencies are separately conserved.
(b) Graphical representation of the (anti)-symmetrized bare interaction vertex
 $\Gamma_0^{ \bar{c} \bar{c} cc \phi \phi} ( 1^{\prime} , 2^{\prime} ; 2  , 1  ; 
k_1 \Omega_1 , k_2 \Omega_2 )$ in Eq.~\eqref{eq:S6freq}.
}
\label{fig:interaction}
\end{figure}

To calculate the disorder average of the  
grand canonical potential and of the correlation functions 
of our model we should use the replica trick \cite{Esterlis19}.
Note that
for the SYK model with Majorana fermions 
replica-nonsymmetric large-$N$ saddle points with energies lower than the
replica-symmetric saddle point have been found~\cite{Arefeva18}.
Whether this happens also for the YSYK model has not been thoroughly investigated.
Here we assume that the  replica symmetry is not broken,
so that we can  simply average the partition function and treat the $g_{ijk}$ as additional complex fields which should be integrated over.
The average partition function can be written as
 \begin{align}
 \langle {\cal{Z}} \rangle 
 = {} & \int {\cal{D}} [ c , \bar{c} , \phi ] e^{ -S_2 } 
 \nonumber
 \\
 & \times \left\langle e^{ -  \frac{1}{\sqrt{NM}}\sum_{ijk} g_{ i j k }  \sum_{\sigma}  
\int_0^{\beta} d \tau   \bar{c}_{i \sigma}  ( \tau )  {c}_{j \sigma} (\tau ) \phi_k 
 ( \tau ) } \right\rangle.
 \label{eq:averagepart}
 \end{align}
Since the probability distribution of the $g_{ijk}$ is Gaussian, the
averaging in Eq.~\eqref{eq:averagepart} generates the usual Debye-Waller factor,
 \begin{equation}
 \langle {\cal{Z}} \rangle  =  \int {\cal{D}} [ c , \bar{c}, \phi ] e^{ -S_2 - S_6 },
 \end{equation} 
where the interaction $S_6$ involves six powers of the fields,
 \begin{align}
 S_6 = {} & - \frac{1}{2NM} \sum_{ijk} \; \sum_{i^\prime j^{\prime} k^{\prime} }  
     \langle g_{ i j k}   g_{ i^{\prime} j^{\prime} k^{\prime} } \rangle 
 \sum_{\sigma \sigma^{\prime}}
 \int_0^{\beta} d \tau \int_0^{\beta} d \tau^{\prime}
 \nonumber
 \\
 & \times  
 \bar{c}_{i \sigma} ( \tau )  {c}_{j \sigma} (\tau ) \phi_k  ( \tau )
\bar{c}_{i^{\prime} \sigma^{\prime}} ( \tau^{\prime} )  {c}_{j^{\prime} \sigma^{\prime}} 
 (\tau^{\prime} ) 
 \phi_{k^{\prime}}  ( \tau^{\prime} ). 
 \label{eq:DebyeWaller}
 \end{align}
Using our assumptions \eqref{eq:secondmom} for the second moments
we obtain
 \begin{widetext}
\begin{align}
 S_6 = {} & - \frac{1}{2 MN } \sum_{ijk} \sum_{ \sigma \sigma^{\prime}}
\int_0^{\beta} d \tau  \bar{c}_{i \sigma} ( \tau )  {c}_{j \sigma} (\tau ) \phi_k  ( \tau ) 
\int_0^{\beta} d \tau^{\prime}
 \Bigl\{ g_1^2 [ 
 \bar{c}_{i \sigma^{\prime}} ( \tau^{\prime} )  {c}_{j \sigma^{\prime}} (\tau^{\prime} ) 
 + ( i \leftrightarrow j)  ]
 \nonumber
 \\
 & \hspace{72mm}
 -  g_2^2 [ 
 \bar{c}_{i \sigma^{\prime}} ( \tau^{\prime} )  {c}_{j \sigma^{\prime}} (\tau^{\prime} ) 
 -  ( i \leftrightarrow j)  ]
\Bigr\} \phi_{k}  ( \tau^{\prime} ) 
 \nonumber
 \\
 = {} &   - \frac{1}{2 MN } \sum_{ijk} \sum_{ \sigma \sigma^{\prime}}
\int_0^{\beta} d \tau \int_0^{\beta} d \tau^{\prime} \phi_k ( \tau ) \phi_k ( \tau^{\prime} ) 
 \Bigl\{ ( g_1^2 - g_2^2 ) 
\bar{c}_{i \sigma} ( \tau )  {c}_{j \sigma} (\tau )  
 \bar{c}_{i \sigma^{\prime}} ( \tau^{\prime} )  {c}_{j \sigma^{\prime}} (\tau^{\prime} )
 \nonumber
 \\
 & \hspace{64mm}
 +   ( g_1^2 + g_2^2 ) 
\bar{c}_{i \sigma} ( \tau )  {c}_{j \sigma} (\tau )  
 \bar{c}_{j \sigma^{\prime}} ( \tau^{\prime} )  {c}_{i \sigma^{\prime}} (\tau^{\prime} )
 \Bigl\}
 \nonumber
 \\ 
 = {} &   - \frac{1}{4 MN } \sum_{ijk} \sum_{ \sigma \sigma^{\prime}}
\int_0^{\beta} d \tau \int_0^{\beta} d \tau^{\prime} \phi_k ( \tau ) \phi_k ( \tau^{\prime} ) 
 \Bigl\{ g_1^2  
 [ \bar{c}_{i \sigma} ( \tau )  {c}_{j \sigma} (\tau ) + ( i \leftrightarrow j ) ]  
 [ \bar{c}_{i \sigma^{\prime}} ( \tau^{\prime} )  {c}_{j \sigma^{\prime}} (\tau^{\prime} )
 + ( i \leftrightarrow j ) ]
 \nonumber
 \\
 &  \hspace{64mm} -  g_2^2  
 [ \bar{c}_{i \sigma} ( \tau )  {c}_{j \sigma} (\tau ) -  ( i \leftrightarrow j ) ]  
 [ \bar{c}_{i \sigma^{\prime}} ( \tau^{\prime} )  {c}_{j \sigma^{\prime}} (\tau^{\prime} )
 -  ( i \leftrightarrow j ) ]
 \Bigl\}.
 \end{align}
In frequency space this interaction  can be written as
 \begin{align}
 S_6 = {}&  -
 \frac{1}{2 N M \beta^6 } \sum_{n_1^{\prime} n_2^{\prime} n_2 n_1  }
 \sum_{ \sigma^{\prime}_1 \sigma_2^{\prime} \sigma_2 \sigma_1}
 \sum_{ \omega_1^{\prime} \omega_2^{\prime} \omega_2 \omega_1} \sum_{ k_1 k_2  }   \sum_{  
 \Omega_1 \Omega_2}
 \beta \delta_{ \omega_1^{\prime} + \omega_2^{\prime} , \omega_2 + \omega_1 + \Omega_2 + \Omega_1 } \beta \delta_{ \omega_2^{\prime} , \omega_2 + \Omega_2 } 
 \delta_{ k_1 , k_2}
 \nonumber
 \\
 & \times \Bigr\{ ( g_1^2 - g_2^2 ) \delta_{ \sigma_1^{\prime} , \sigma_1 }
 \delta_{ \sigma_2^{\prime} , \sigma_2 } \delta_{ n_1^{\prime} , n_2^{\prime}}
 \delta_{ n_1 , n_2 } 
  + ( g_1^2 + g_2^2 ) \delta_{ \sigma_1^{\prime} , \sigma_1 }
 \delta_{ \sigma_2^{\prime} , \sigma_2 } \delta_{ n_1^{\prime} , n_2}
 \delta_{ n_2^{\prime} , n_1 } 
 \Bigr\} 
 \bar{c}_{ 1^{\prime}} \bar{c}_{ 2^{\prime} }
 c_{2 } c_{1} \phi_{ k_1 \Omega_1 } \phi_{ k_2 \Omega_2},
 \label{eq:S6freq}
 \end{align}
\end{widetext}
where we have introduced the abbreviations $1 = ( n_1 , \sigma_1 , \omega_1 )$,
$1^{\prime} = ( n_1^{\prime} , \sigma_1^{\prime} , \omega_1^{\prime} )$ 
etc.~for the fermionic labels.
A graphical representation of the non-symmetrized interaction vertices
in Eq.~\eqref{eq:S6freq} is shown in Fig.~\ref{fig:interaction}.
For the derivation of formally exact FRG flow equations it is convenient 
to work with vertices which are antisymmetric
with respect to permutation of 
the  outgoing and the  incoming fermion labels, and symmetric with respect to
permutation of the boson labels. 
Therefore we write the bare interaction
$S_6$ in the symmetrized form
 \begin{align}
 & S_6  =  \frac{1}{(2!)^3 N M \beta^6 } \sum_{n_1^{\prime} n_2^{\prime} n_2 n_1  }
 \sum_{ \sigma^{\prime}_1 \sigma_2^{\prime} \sigma_2 \sigma_1}
 \sum_{ \omega_1^{\prime} \omega_2^{\prime} \omega_2 \omega_1} \sum_{ k_1 k_2  }  \sum_{  
 \Omega_1 \Omega_2}
 \nonumber
 \\
 & \times 
 \beta \delta_{ \omega_1^{\prime} + \omega_2^{\prime} , \omega_2 + \omega_1 + \Omega_1 + \Omega_2 }
 \Gamma_0^{ \bar{c} \bar{c} cc \phi \phi} ( 1^{\prime} , 2^{\prime} ; 2  , 1  ; 
k_1 \Omega_1 , k_2 \Omega_2 )
 \nonumber
 \\
 & \times  \bar{c}_{ 1^{\prime}} \bar{c}_{ 2^{\prime} }
 c_{2 } c_{1} \phi_{ k_1 \Omega_1 } \phi_{ k_2 \Omega_2},
 \label{eq:S6freq}
 \end{align}
where the mixed six-point vertex 
 $\Gamma_0^{ \bar{c} \bar{c} cc \phi \phi} ( 1^{\prime} , 2^{\prime} ; 2  , 1  ; k_1 \Omega_1 , k_2 \Omega_2)$ is antisymmetric with respect to the exchange $ 1^{\prime} \leftrightarrow 2^{\prime}$ and $1 \leftrightarrow 2$, and symmetric with respect to the exchange $k_1 \Omega_1
 \leftrightarrow k_2 \Omega_2$. 
 Explicitly, the bare value of the properly symmetrized 
mixed six-point vertex is
 \begin{widetext}
 \begin{align}
 &  \Gamma_0^{ \bar{c} \bar{c} cc \phi \phi} ( n_1^{\prime} \sigma_1^{\prime} \omega_1^{\prime}
, n_2^{\prime} \sigma_2^{\prime} \omega_2^{\prime}; n_2 \sigma_2 \omega_2  , 
 n_1 \sigma_1 \omega_1  ; k_1 \Omega_1 , k_2  \Omega_2 ) = - \delta_{k_1  k_2} \frac{\beta}{2} \Biggl\{
\nonumber
 \\
 & 
( g_1^2 + g_2^2 ) \bigl[ \delta_{ n_1^{\prime}  n_2} \delta_{ n_2^{\prime}  n_1 }
 \delta_{\sigma_1^{\prime}  \sigma_1 } \delta_{\sigma_2^{\prime}  \sigma_2} ( \delta_{ \omega_1^{\prime} , \omega_1 + \Omega_1 } + 
  \delta_{ \omega_2^{\prime} , \omega_2 + \Omega_2 })
 - \delta_{ n_2^{\prime}  n_2} \delta_{ n_1^{\prime}  n_1 }
\delta_{\sigma_2^{\prime}  \sigma_1 } \delta_{\sigma_1^{\prime}  \sigma_2} ( \delta_{ \omega_2^{\prime} , \omega_1 + \Omega_1 } + 
  \delta_{ \omega_1^{\prime} , \omega_2 + \Omega_2 })
\bigr] 
 \nonumber
 \\
 & + 
  ( g_1^2 - g_2^2 ) \delta_{ n_1^{\prime}  n_2^{\prime}} \delta_{n_1  n_2}
\bigl[ \delta_{\sigma_1^{\prime}  \sigma_1 } \delta_{\sigma_2^{\prime}  \sigma_2} ( \delta_{ \omega_1^{\prime} , \omega_1 + \Omega_1 } + 
  \delta_{ \omega_2^{\prime} , \omega_2 + \Omega_2 })
 -
\delta_{\sigma_2^{\prime}  \sigma_1 } \delta_{\sigma_1^{\prime}  \sigma_2} ( \delta_{ \omega_2^{\prime} , \omega_1 + \Omega_1 } + 
  \delta_{ \omega_1^{\prime} , \omega_2 + \Omega_2 })
\bigr]  
+ ( \Omega_1 \leftrightarrow \Omega_2 )
 \Biggr\}.
 \label{eq:sixpointsym}
 \end{align}
For later reference let us also calculate the site-averaged mixed six-point vertex
 \begin{equation}
\Gamma_0^{ \bar{c} \bar{c} cc \phi \phi} (  \sigma_1^{\prime} \omega_1^{\prime}
,  \sigma_2^{\prime} \omega_2^{\prime};  \sigma_2 \omega_2  , 
 \sigma_1 \omega_1  ;  \Omega_1 ,  \Omega_2 ) = 
\frac{1}{N^2 M}
 \sum_{ n_1 n_2 =1 }^N \sum_{ k =1}^M
 \Gamma_0^{ \bar{c} \bar{c} cc \phi \phi} ( n_1 \sigma_1^{\prime} \omega_1^{\prime}
, n_2 \sigma_2^{\prime} \omega_2^{\prime}; n_2 \sigma_2 \omega_2  , 
 n_1 \sigma_1 \omega_1  ; k \Omega_1 , k  \Omega_2 ).
 \label{eq:sixpointaverage}
 \end{equation}
Actually, for our purpose we only need this vertex in the limit of large $N$ and $M$ where
only the terms involving $\delta_{ n_1^{\prime} , n_1 } \delta_{ n_2^{\prime} , n_2 }$
in Eq.~\eqref{eq:sixpointsym} contribute,
 \begin{equation}
   \Gamma_0^{ \bar{c} \bar{c} cc \phi \phi} (  \sigma_1^{\prime} \omega_1^{\prime}
,  \sigma_2^{\prime} \omega_2^{\prime};  \sigma_2 \omega_2  , 
 \sigma_1 \omega_1  ;  \Omega_1 ,  \Omega_2 ) = \frac{\beta}{2}
  (g_1^2 + g_2^2)  \delta_{\sigma_2^{\prime} , \sigma_1 } \delta_{\sigma_1^{\prime} , \sigma_2} ( \delta_{ \omega_2^{\prime} , \omega_1 + \Omega_1 } + 
  \delta_{ \omega_1^{\prime} , \omega_2 + \Omega_2 })  + ( \Omega_1 \leftrightarrow \Omega_2 )
 + {\cal{O}} ( 1 / N ).
 \label{eq:sixpointinit}
 \end{equation}
\end{widetext}

To derive formally exact FRG flow equations for the irreducible vertices of the model  
defined by the bare action $S_2 + S_6$  
in Eqs.~(\ref{eq:S2def}, \ref{eq:S6freq})
we add frequency-dependent regulators $R_{\Lambda} ( \omega )$ and $R^{\phi}_{\Lambda} ( \Omega )$
to the quadratic 
part of the action, where the cutoff-parameter $\Lambda$ should be chosen such
that the regulators vanish for $\Lambda =0$ and diverge for $\Lambda \rightarrow \infty$.
At this point it is not necessary to specify the regulators.
The modified quadratic part of the Euclidean action is then
  \begin{align}
 S_{2 , \Lambda} 
 = {} & - \frac{1}{\beta} \sum_{n \sigma  \omega }  G^{-1}_{0, \Lambda} ( \omega )   
 \bar{c}_{ n \sigma \omega } c_{ n \sigma  \omega }
 \nonumber
 \\
 & + \frac{1}{2 \beta} \sum_{k \Omega} F^{-1}_{ 0 , \Lambda} ( \Omega ) 
 \phi^{\ast}_{k  \Omega} \phi_{k \Omega},
 \end{align}
where we have introduced the regularized inverse propagators
 \begin{align}
 G^{-1}_{0, \Lambda} ( \omega )   & =  i \omega + \mu - R_{\Lambda} ( \omega ),
 \\
  F^{-1}_{0, \Lambda} ( \Omega )   & =  | \Omega |  + \Delta +  R^{\phi}_{\Lambda} ( \Omega ).
 \end{align}
To obtain the flow equation for the
generating functional of the irreducible vertices of the YSYK model
for finite $N$ and $M$ we introduce  
the cutoff-dependent generating functional
of the connected imaginary time correlation functions,
 \begin{equation}
 e^{ {\cal{G}}_\Lambda [ \bar{j} , j , J ] } = \frac{ \int  {\cal{D}}  [ \bar{c} , c , \phi ] e^{ -S_{2, \Lambda} - S_6 + 
( \bar{j} , c ) + ( \bar{c} , j ) + ( J , \phi) } }{  \int  {\cal{D}}  [ \bar{c} , c , \phi] e^{ -S_{2, \Lambda} }},
 \end{equation}
where $\bar{j}_{ n \sigma \omega}$ and $j_{ n \sigma \omega }$ are independent 
Grassmann sources, $J_{ k \Omega}$ is a bosonic source,
 and we have introduced the notations
 \begin{align}
( \bar{j} , c ) + ( \bar{c} , j ) & =  \frac{ 1}{\beta} \sum_{n \sigma \omega} [ \bar{j}_{ n \sigma \omega} c_{ n \sigma \omega }
 + \bar{c}_{ n \sigma \omega} j_{ n \sigma \omega } ],
 \\
 ( J , \phi ) & = \frac{1}{\beta} \sum_{ k \Omega} J^{\ast}_{ k \Omega} \phi_{ k \Omega}.
 \end{align}
The generating functional of the irreducible vertices (average effective action) 
can  now be defined via the
subtracted Legendre transform
 \begin{align}
& \Gamma_{\Lambda} [ \langle \bar{c} \rangle , \langle c \rangle, 
\langle \phi \rangle   ]  
\nonumber\\
& =  ( \bar{j} , \langle c \rangle  ) + ( \langle \bar{c} \rangle , j ) 
 + ( J , \langle \phi \rangle )
-  
 {\cal{G}}_\Lambda [ \bar{j} , j , J ]  
 \nonumber
 \\
 &   
- \frac{1}{\beta} \sum_{ n \sigma  \omega} R_{\Lambda} ( \omega ) \langle 
 \bar{c}_{ n \sigma \omega}\rangle  \langle c_{ n \sigma \omega } \rangle
- \frac{1}{2 \beta} \sum_{k \Omega} R^{\phi}_{\Lambda} ( \Omega ) 
 \langle \phi^\ast_{ k  \Omega } \rangle \langle \phi_{ k \Omega } \rangle,
 \end{align}
where on the right-hand side the sources $\bar{j}$ and $j$ 
should be expressed in terms of the source-dependent expectation values $\langle \bar{c} \rangle $,
$\langle c \rangle$, and $\langle \phi \rangle$  by inverting the relations
 \begin{subequations}
 \begin{align}
  \langle c_{ n \sigma \omega} \rangle & = 
 \frac{ \delta {\cal{G}}_{\Lambda} [ \bar{j} , j , J ] }{ \delta \bar{j}_{ n \sigma \omega }},
 \\
 \langle \bar{c}_{ n \sigma \omega} \rangle & =  
 - \frac{ \delta {\cal{G}}_{\Lambda} [ \bar{j} , j , J  ] }{ \delta {j}_{ n \sigma \omega }},
 \\
 \langle \phi_{ k \Omega}  \rangle & =  
\frac{ \delta {\cal{G}}_{\Lambda} [ \bar{j} , j , J ] }{ \delta {J}^{\ast}_{ k \Omega }}.
 \end{align}
 \end{subequations}
The functional $   \Gamma_{\Lambda} [ \langle \bar{c} \rangle , \langle c \rangle, 
\langle \phi \rangle   ]$ satisfies the usual Wetterich equation \cite{Wetterich93,Berges2002,Metzner2012,Dupuis2021}  for systems involving  both bosonic and fermionic fields \cite{Kopietz10}.
For notational simplicity we will now rename $\langle \bar{c} \rangle \rightarrow
 \bar{c}$, $\langle c \rangle \rightarrow c$ and $\langle \phi \rangle \rightarrow \phi$, 
 i.e., the symbols $\bar{c}$, $c$ and $\phi$ now denote the expectation values of the corresponding quantum fields in the presence of sources.
For finite $N$ and $M$ the first few terms in the vertex expansion of the functional
$\Gamma_{\Lambda} [ \bar{c} , c, \phi ]$ then have the following structure,
 \begin{widetext}
 \begin{align}
 &  \Gamma_{\Lambda} [ \bar{c} , c, \phi ]  =   \beta \Omega_{\Lambda}
 +  \frac{1}{\beta} \sum_{ n^{\prime} n  \sigma \omega}
 [  - \delta_{ n n^{\prime}}(  i \omega + \mu) +  \Sigma_{\Lambda}^{ 
n^{\prime} n}   (  \omega ) ]
 \bar{c}_{ n^{\prime}  \sigma \omega } c_{n  \sigma \omega } 
 + \frac{1}{2 \beta} \sum_{ k^{\prime} k \Omega} [
 \delta_{ k  k^{\prime}} (
 | \Omega | + \Delta )  + \Pi^{k^{\prime} k }_{ \Lambda} ( \Omega ) ]
 \phi^{\ast}_{ k^{\prime} \Omega} \phi_{ k \Omega }
 \nonumber
 \\ 
 & +
\frac{1}{  (2!)^2    N \beta^4} \sum_{ n_1^{\prime} n_2^{\prime} n_2 n_1 }  
 \sum_{\sigma_1^{\prime} \sigma_2^\prime \sigma_2 \sigma_1} 
 \sum_{ \omega_1^{\prime} 
 \omega_2^{\prime} \omega_2 \omega_1 } \beta \delta_{ \omega_1^{\prime} + \omega_2^{\prime} ,
  \omega_2 + \omega_1 } 
 \Gamma_{\Lambda}^{\bar{c} \bar{c} cc} ( 1^{\prime},
 2^{\prime}  ; 2  , 1 )
\bar{c}_{ n_1^{\prime} \sigma_1^{\prime} \omega_1^{\prime}} \bar{c}_{ n_2^{\prime} 
 \sigma_2^{\prime} \omega_2^{\prime}}
 c_{ n_2 \sigma_2 \omega_2 } c_{ n_1 \sigma_1\omega_1 } 
 \nonumber
 \\
 &  +
\frac{1}{  4!   M \beta^4} \sum_{ k_1 k_2 k_3 k_4 }  
\sum_{ \Omega_1 \ldots \Omega_4}
 \beta \delta_{ \Omega_1 + \Omega_2 +
 \Omega_3 + \Omega_4,0 } 
 \Gamma_{\Lambda}^{\phi \phi \phi \phi } ( k_1 \Omega_1 ,
 k_2  \Omega_2 ,  k_3 \Omega_3 , k_4 \Omega_4 )
 {\phi}_{ k_1 \Omega_1} {\phi}_{ k_2 \Omega_2}
 \phi_{ k_2 \Omega_3 } \phi_{ n_4 \Omega_4 } 
 \nonumber
 \\
 & +  
\frac{1}{  2!    M \beta^4} \sum_{ n^{\prime}  n k_1 k_2 }  \sum_{\sigma}
\sum_{ \omega^{\prime} 
 \omega \Omega_1 \Omega_2 } \beta \delta_{ \omega^{\prime},  \omega +
 \Omega_1 + \Omega_2 } 
 \Gamma_{\Lambda}^{ \bar{c} c \phi \phi} ( n^{\prime} \sigma \omega^{\prime} ; n \sigma \omega; k_1 \Omega_1
 k_2 \Omega_2 ) \bar{c}_{ n^{\prime}  \sigma \omega^{\prime} } c_{ n \sigma \omega} \phi_{ k_1 \Omega_1 }
 \phi_{ k_2 \Omega_2 }
 \nonumber
 \\
 & +  \frac{1}{(2!)^3 N M \beta^6 } \sum_{n_1^{\prime} n_2^{\prime} n_2 n_1  }
 \sum_{ \sigma^{\prime}_1 \sigma_2^{\prime} \sigma_2 \sigma_1}
 \sum_{ \omega_1^{\prime} \omega_2^{\prime} \omega_2 \omega_1} 
 \sum_{ k_1 k_2  }  \sum_{  
 \Omega_1 \Omega_2} 
 \beta \delta_{ \omega_1^{\prime} + \omega_2^{\prime} , \omega_2 + \omega_1 + \Omega_1 + \Omega_2 }
 \Gamma_\Lambda^{ \bar{c} \bar{c} cc \phi \phi} ( 1^{\prime} , 2^{\prime} ; 2  , 1  ; 
k_1 \Omega_1 , k_2 \Omega_2 )
 \nonumber
 \\
 & \hspace{75mm} \times  \bar{c}_{ n_1^{\prime} \sigma_1^{\prime} \omega_1^{\prime} } \bar{c}_{ n_2^{\prime} \sigma_2^{\prime} \omega_2^{\prime}}
 c_{n_2 \sigma_2 \omega_2 } c_{n_1 \sigma_1 \omega_1} \phi_{ k_1 \Omega_1 } \phi_{ k_2 \Omega_2},
+ \ldots \; ,
 \label{eq:vertexexp}
 \end{align}
 \end{widetext}
where in the arguments of the fermionic four-point vertex and in the six-point vertex
we have abbreviated $  n_1 \sigma_1 \omega_1 \rightarrow 1$ and similarly for the
other fermionic labels, and 
we have omitted other six-point vertices involving different field combinations as well as
vertices involving more than six fields.

\begin{figure}[tb]
 \begin{center}
  \centering
\vspace{7mm}
 \includegraphics[width=0.45\textwidth]{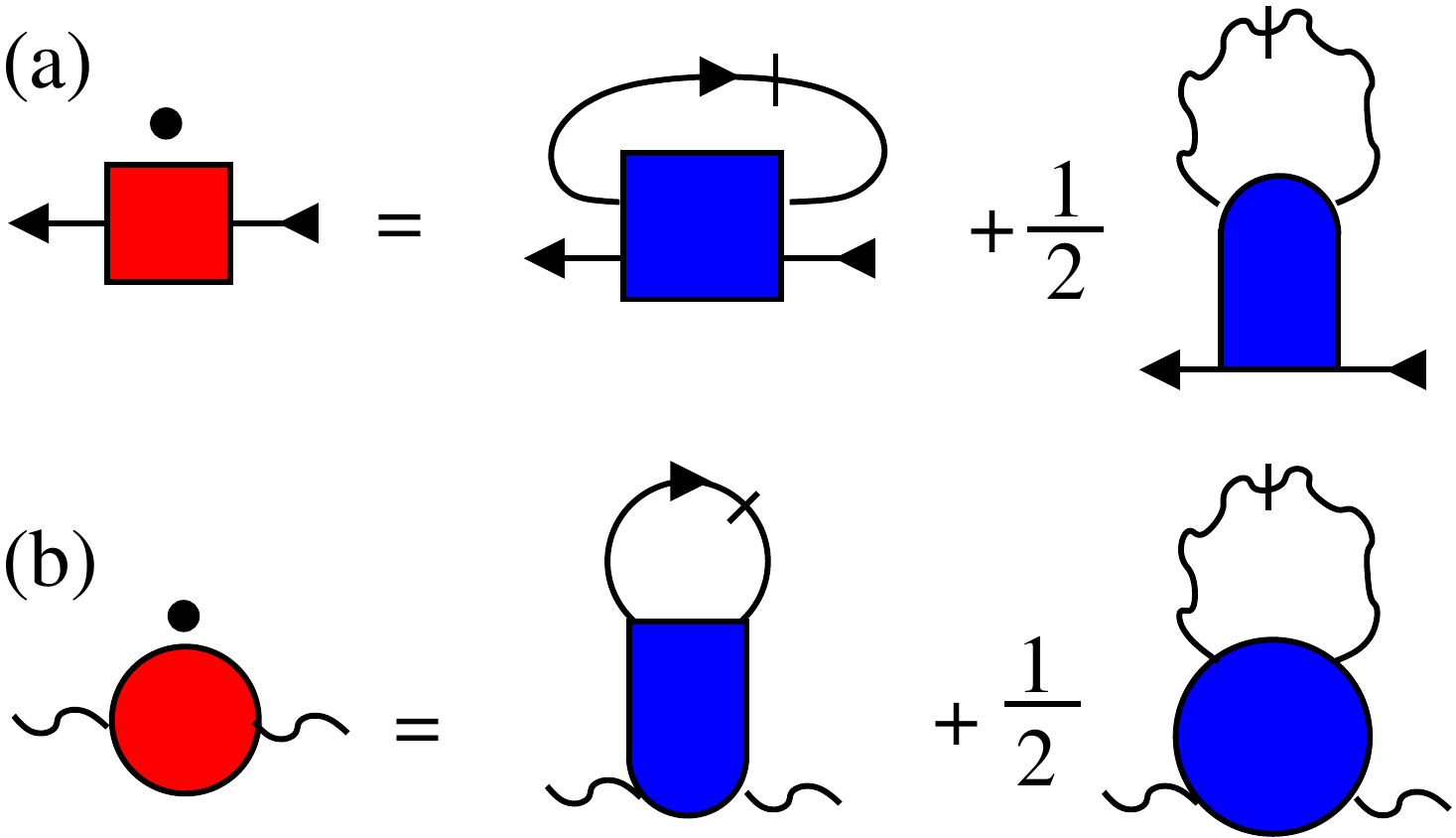}
   \end{center}
  \caption{Graphical representation of the exact FRG flow equations for the irreducible two-point vertices:
  (a) flow equation \eqref{eq:flowself} of the 
fermionic self-energy; (b) flow equation \eqref{eq:flowPi}
of the
 bosonic self-energy. The dot above the vertices on the left-hand side
represents the scale derivative $\partial_{\Lambda}$.
Outgoing arrows represent external legs associated with $\bar{c}$ 
while incoming arrows represent $c$. 
The external wavy lines
correspond to $\phi$.
Slashed lines with arrows represent the fermion
single-scale propagator $\dot{G}_{\Lambda} ( \omega )$,
while slashed wavy lines represent the boson single-scale propagator
$\dot{F}_{\Lambda} ( \Omega )$.
The blue symbols represent three different types of four-point vertices distinguished by the externals legs.
}
\label{fig:flow2point}
\end{figure}

\begin{figure}[tb]
 \begin{center}
  \centering
\vspace{7mm}
 \includegraphics[width=0.45\textwidth]{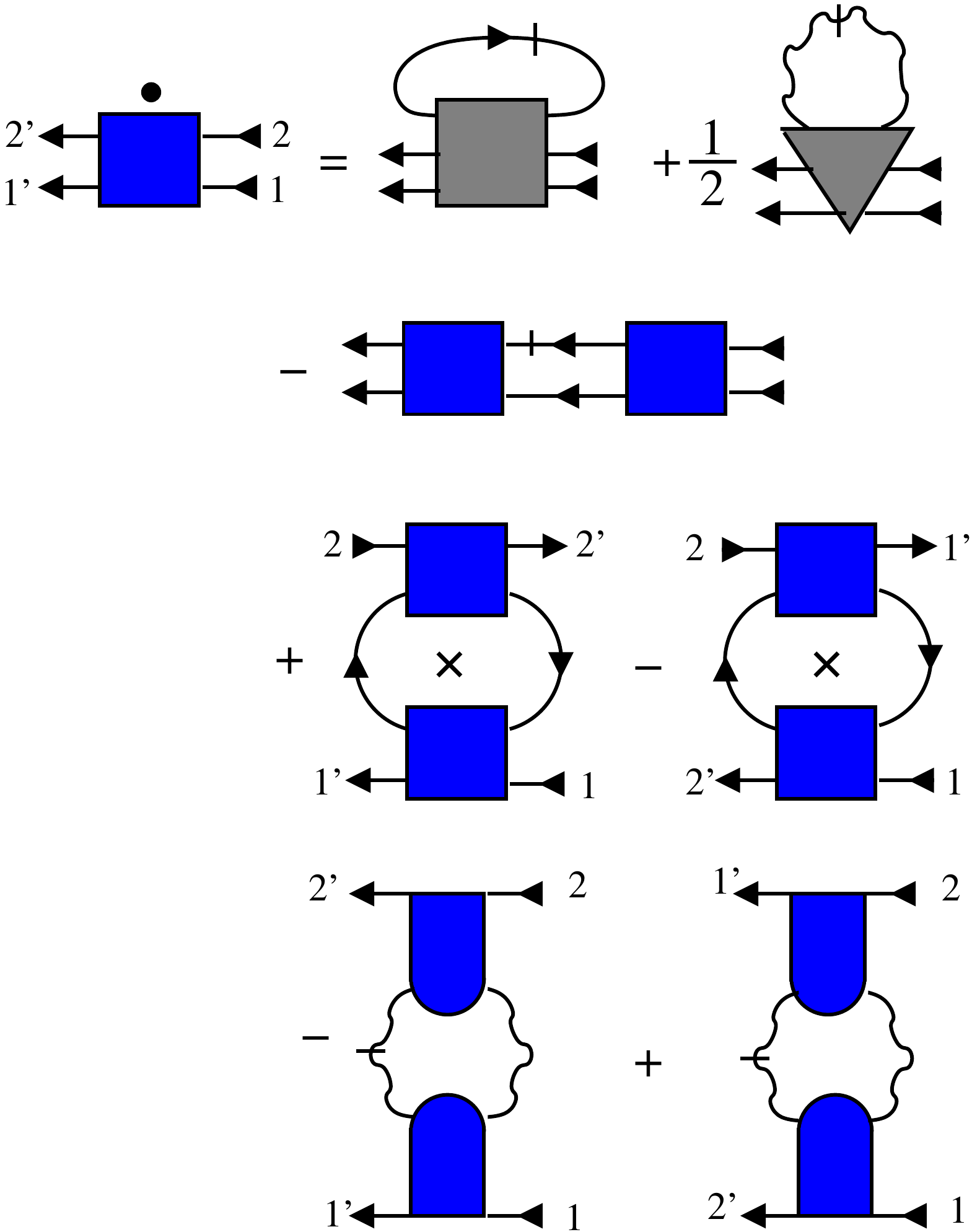}
   \end{center}
  \caption{Graphical representation of the exact FRG flow equation \eqref{eq:flow4point}
  for the fermionic $4$-point vertex.
The gray square represents the fermionic six-point vertex, while the gray triangle represents
the mixed six-point vertex with four fermionic and two bosonic external legs.
The internal lines with arrows represent the exact scale-dependent fermion propagator
$G_{\Lambda} ( \omega )$,
whereas
wavy internal lines represent the boson propagator $F_{\Lambda} ( \Omega )$.
The cross inside the loops means that one should sum up two diagrams where 
either the right or the left  propagator forming the loop is replaced by the corresponding single-scale propagator. 
The rest of the symbols is defined in the caption of Fig.~\ref{fig:flow2point}.
}
\label{fig:flow4point}
\end{figure}

Using the general flow equations generated by the vertex expansion
of the Wetterich equation given in Ref.~[\onlinecite{Kopietz10}], we may 
now write down formally exact flow equations for vertices in the above expansion for finite
$N$ and $M$.
The scale-dependent grand canonical potential satisfies the exact flow equation
 \begin{align}
  \partial_{\Lambda} \Omega_{\Lambda} 
   = {} &  \frac{1}{\beta} \sum_{n \sigma \omega} {G}^{nn}_{\Lambda} ( \omega  )  \partial_{\Lambda}
    R_{\Lambda} ( \omega )
   \nonumber
   \\
   & + \frac{1}{2 \beta} \sum_{k  \Omega} {F}_{\Lambda}^{kk} ( \Omega  )  \partial_{\Lambda}
    R_{\Lambda}^{\phi} ( \Omega ),
   \end{align}
 where  ${G}^{nn}_{\Lambda} ( \omega  )$ and 
 ${F}^{kk}_{\Lambda} ( \Omega  )$ are the diagonal elements of the
 scale-dependent  propagator matrices whose inverse is given by the following matrices in the 
 flavor-indices,
 \begin{align}
  {[} {\mathbf{G}}^{-1}_{\Lambda} ( \omega )  {]}^{n n^{\prime}} & = \delta_{ n n^{\prime} } 
   [ i \omega + \mu - R_{\Lambda} ( \omega ) ] - \Sigma^{ n  n^{\prime}}_{\Lambda} ( \omega ),
   \phantom{a}
   \\
   {[} \mathbf{F}^{-1}_{\Lambda} ( \Omega ) {]}^{k k^{\prime}} & = \delta_{ k k^\prime}
   [ | \Omega | + \Delta + R_{\Lambda}^{\phi} ( \Omega ) ] + \Pi^{k k^\prime}_{\Lambda} ( \Omega ).
 \end{align}
The irreducible fermionic self-energy $\Sigma^{ n n^{\prime}}_{\Lambda} ( \omega )$
satisfies the exact flow equation
 \begin{align}
 &  \partial_{\Lambda} \Sigma^{n n^{\prime} }_{\Lambda} (  \omega ) =
  \nonumber
  \\
  &
 \frac{1}{N \beta} \sum_{  m  m^{\prime} } 
 \sum_{ \sigma^{\prime} \omega^{\prime} } \dot{G}^{ m m^{\prime} }_{\Lambda} (   \omega^{\prime} ) 
\Gamma^{\bar{c} \bar{c} cc}_{\Lambda} ( n \sigma \omega , m  \sigma^{\prime} \omega^{\prime}  ; m^{\prime} 
 \sigma^{\prime} \omega^{\prime} , n^{\prime} \sigma  \omega )
 \nonumber
 \\
 & +\frac{1}{2 M \beta} \sum_{  k  k^{\prime}} 
 \sum_{ \Omega } \dot{F}^{  k k^{\prime} }_{\Lambda} (  \Omega ) 
\Gamma^{\bar{c} c \phi \phi }_{\Lambda} ( n  \sigma \omega ; n^{\prime}  \sigma \omega ; k \bar{\Omega} , k^{\prime} \Omega ),
 \label{eq:flowself}
 \end{align}
while the scale-dependent bosonic self-energy satisfies
 \begin{align}
 &  \partial_{\Lambda} \Pi^{k k^{\prime} }_{\Lambda} (  \Omega )  = 
 \nonumber
 \\ 
 & 
 \frac{1}{M \beta} \sum_{  n   n^{\prime} } 
 \sum_{ \sigma \omega } \dot{G}^{ n  n^{\prime} }_{\Lambda} (   \omega ) 
\Gamma^{\bar{c} {c} \phi \phi}_{\Lambda} ( n  \sigma  \omega ; n^{\prime}  \sigma \omega  
 ; k \bar{\Omega},   k^{\prime}  \Omega )
 \nonumber
 \\
 &   \hspace{-3mm} +  \frac{1}{2 M \beta} \sum_{  l l^{\prime}} 
 \sum_{ \Omega^{\prime} } \dot{F}^{  l  l^{\prime} }_{\Lambda} (  \Omega^{\prime} ) 
\Gamma^{\phi \phi \phi \phi }_{\Lambda} ( l   \bar{\Omega}^{\prime} , l^{\prime} \Omega^{\prime}, k \bar{ \Omega} ,  k^{\prime} \Omega ),
 \label{eq:flowPi}
 \end{align}
where we have introduced the abbreviation $\bar{\Omega} = - \Omega$ and
 $\dot{G}^{n n^{\prime}}_{\Lambda} (  \omega )$ and
$\dot{F}^{k k^{\prime}}_{\Lambda} ( \Omega )$ 
are the fermionic and bosonic single-scale propagators \cite{Kopietz10}.
Graphical representations of the 
flow equations \eqref{eq:flowself} and \eqref{eq:flowPi}
are shown in Fig.~\ref{fig:flow2point}.
The right-hand sides of these flow equations
depend on three different types of four-point vertices, 
which in turn satisfy flow equations involving not only four-point vertices 
but also various types of six-point vertices. 
For example, the exact flow equation for the fermionic four-point vertex is 
 \begin{align}
   & \partial_{\Lambda} \Gamma^{\bar{c} \bar{c} cc}_{\Lambda} ( 
 n_1^{\prime}   \sigma_1^{\prime}  \omega_1^{\prime} , n_2^{\prime}  \sigma_2^{\prime} \omega_2^{\prime}  ; n_2 \sigma_2 
 \omega_2 , n_1  \sigma_1 \omega_1 )
 \nonumber
 \\
 = {} & 
 \frac{1}{N \beta } \sum_{ n n^{\prime}  }\sum_{\sigma \omega}
 \dot{G}^{n n^{\prime} }_{\Lambda} ( \omega )
 \Gamma^{ \bar{c} \bar{c} \bar{c} ccc}_{\Lambda} (
 1^{\prime} , 2^{\prime} , n  \sigma \omega  ; n^{\prime} \sigma \omega , 2, 1 )
 \nonumber
 \\
 & +   \frac{1}{2 M \beta} \sum_{ k k^{\prime}} \sum_{\Omega}
 \dot{F}^{k k^{\prime}}_{\Lambda} ( \Omega )
\Gamma^{ \bar{c} \bar{c} cc \phi \phi}_{\Lambda} (
 1^{\prime} , 2^{\prime} ; 2 , 1 ; k \bar{\Omega} , k^{\prime} \Omega)
 \nonumber
 \\
 & +   L^{\rm pp}_{\Lambda} ( 1^{\prime} , 2^{\prime} ; 2 , 1 ) +  L^{\rm ph}_{\Lambda}    ( 1^{\prime} , 2^{\prime} ; 2 , 1 )
 +
L^{\rm bos}_{\Lambda} (  1^{\prime} , 2^{\prime} ; 2 , 1   ),
 \label{eq:flow4point}
 \end{align}
where on the right-hand side we have used again the abbreviation
$1 = ( n_1 \sigma_1 \omega_1)$, and in the last line we have introduced three
different loop contributions,
 \begin{widetext}
 \begin{align}
L^{\rm pp}_{\Lambda} ( 1^{\prime} , 2^{\prime} ; 2 , 1 ) 
 = {} & - \frac{1}{ N^2 \beta}
 \sum_{ n n^{\prime}} \sum_{ m  m^{\prime }  } \sum_{ \sigma \sigma^{\prime} }
 \sum_{ \omega} 
 \dot{G}^{n n^{\prime} }_{\Lambda} ( \omega ) G_{\Lambda}^{ m m^{\prime}  } ( \omega_1 + \omega_2 - \omega ) 
 \nonumber
 \\
 & \times 
 \Gamma^{ \bar{c} \bar{c} cc }_{\Lambda} ( 1^{\prime} , 2^{\prime} ; n \sigma^{\prime} \omega_1 + \omega_2 - \omega, m \sigma \omega )
 \Gamma^{ \bar{c} \bar{c} cc }_{\Lambda} ( m^{\prime} \sigma \omega, 
 n^{\prime} \sigma^{\prime} \omega_1 + \omega_2 - \omega;  2 , 1 ),
 \label{eq:Lpp}
 \\
L^{\rm ph}_{\Lambda} ( 1^{\prime} , 2^{\prime} ; 2 , 1 ) 
 = {} &  \frac{1}{ N^2 \beta}
 \sum_{ n n^{\prime}} \sum_{ m  m^{\prime }  } \sum_{ \sigma \sigma^{\prime} }
 \sum_{ \omega} 
 \left[
 \dot{G}^{n n^{\prime} }_{\Lambda} ( \omega ) 
 G_{\Lambda}^{ m m^{\prime}  } ( \omega + \omega_1  - \omega_1^{\prime}  ) 
 + 
 {G}^{n n^{\prime} }_{\Lambda} ( \omega ) 
 \dot{G}_{\Lambda}^{ m m^{\prime}  } ( \omega + \omega_1  - \omega_1^{\prime}  )  
  \right]
 \nonumber
 \\
 & \times 
 \Gamma^{ \bar{c} \bar{c} cc }_{\Lambda} ( 1^{\prime} , m^{\prime} \sigma^{\prime} \omega + \omega_1 - \omega_1^{\prime} ; n \sigma \omega, 1 )
 \Gamma^{ \bar{c} \bar{c} cc }_{\Lambda} (  2^{\prime}, n^{\prime} \sigma \omega; 
 m \sigma^{\prime} \omega + \omega_1 - \omega_1^{\prime} ,   2  )
 \nonumber
 \\
 & 
- \left\{  ( n_1 \sigma_1 \omega_1 ) \leftrightarrow  ( n_2 \sigma_2 \omega_2 ) \right\}
,
 \label{eq:Lph}
 \\
L^{\rm bos}_{\Lambda} ( 1^{\prime} , 2^{\prime} ; 2 , 1 ) 
= {} &  - \frac{1}{ M^2 \beta}
 \sum_{ k k^{\prime}} \sum_{ l  l^{\prime }  } 
 \sum_{ \Omega} 
 \dot{F}^{k k^{\prime} }_{\Lambda} ( \Omega ) 
 F_{\Lambda}^{ l l^{\prime}  } ( \Omega + \omega_1  - \omega_1^{\prime}  ) 
 \nonumber
 \\
 & \times 
 \Gamma^{ \bar{c}{c} \phi \phi }_{\Lambda} ( 1^{\prime} ; 1 ; k \Omega, l \bar{\Omega} - \omega_1 + \omega_1^{\prime}  )
 \Gamma^{ \bar{c} c \phi \phi }_{\Lambda} (  2^{\prime}; 2 ; k^{\prime} \bar{\Omega} ,
l^{\prime} \Omega - \omega_2 + \omega_2^{\prime} )
 \nonumber
 \\
 &
+ \left\{  ( n_1 \sigma_1 \omega_1 ) \leftrightarrow  ( n_2 \sigma_2 \omega_2 ) \right\} .
 \label{eq:Lbos}
 \end{align}
 \end{widetext}
A graphical representation of the exact flow equation \eqref{eq:flow4point}
for the fermionic four-point vertex 
is shown in Fig.~\ref{fig:flow4point}.
The six-point vertices on the right-hand side of Eq.~\eqref{eq:flow4point}
satisfy flow equations involving various types of eight-point vertices.
Obviously, for finite $N$ and $M$ the FRG vertex expansion generates an infinite
hierarchy of flow equations.

The crucial point is now that in the limit $N \rightarrow \infty$ and
$M \rightarrow \infty$ the system of flow equations can be closed 
to leading order in $1/N$ and $1/M$ where we can retain only the 
contributions to  the flow equations which have  finite limits for $N \rightarrow 
\infty$ and $M \rightarrow \infty$.
The self-energies are then diagonal in the site-indices,
 \begin{align}
  \Sigma^{n n^\prime}_{\Lambda} ( \omega ) & = \delta_{n n^\prime} \Sigma_{\Lambda} ( \omega )
 + {\cal{O}} ( 1/N ),
  \\  
  \Pi^{kk^\prime}_{\Lambda} ( \Omega ) & = \delta_{kk^\prime} \Pi_{\Lambda} ( \Omega ) 
  + {\cal{O}} ( 1/N) 
  ,
  \end{align}
where the symbol ${\cal{O}} (1/N)$ represents also terms of order $1/M$ because we assume
that  the ratio $N/M$ is of order unity.
The leading large-$N$ limit of the four-point vertices is more complicated.
Consider first the fermionic four-point vertex
$\Gamma^{\bar{c} \bar{c} cc}_{\Lambda} ( n_1^{\prime} \sigma_1^{\prime} \omega_1^{\prime} , n_2^{\prime} \sigma_2^{\prime} \omega_2^{\prime} ;
 n_2 \sigma_2 \omega_2 , n_1 \sigma \omega_1 ) $. We obtain a consistent 
large-$N$ truncation of the hierarchy of FRG flow equations if we assume that 
the leading components of the fermionic four-point vertex are constrained
by the condition that
 site labels $ n_1^{\prime}$ and $n_2^\prime$ of the two outgoing fermions agree with 
 the site-labels $n_1$ and $n_2$ of the two incoming fermions up to a permutation.
Taking into account the antisymmetry of
$\Gamma^{\bar{c} \bar{c} cc}_{\Lambda} ( 1^{\prime} , 2^{\prime}; 2 , 1 ) $ with respect to the exchange $1^{\prime} \leftrightarrow 2^{\prime}$ and
 $1 \leftrightarrow 2$, this implies
 \begin{align}
& \Gamma^{\bar{c} \bar{c} c c }_{\Lambda} ( n_1^{\prime} \sigma_1^{\prime} \omega_1^{\prime} , n_2^{\prime} \sigma_2^{\prime} \omega_2^{\prime} ;
 n_2 \sigma_2 \omega_2 , n_1 \sigma_1 \omega_1 ) =
 \nonumber
 \\
&    \delta_{ n_1^{\prime} n_1 } \delta_{ n_2^{\prime} n_2 }
  \Gamma^{\bar{c} \bar{c} cc}_{\Lambda} ( \sigma_1^{\prime} \omega_1^{\prime} , 
  \sigma_2^{\prime} \omega_2^{\prime} ;
  \sigma_2 \omega_2 ,  \sigma_1 \omega_1 ) 
 \nonumber
 \\
 & - \delta_{ n_1^{\prime} n_2 } \delta_{ n_2^{\prime} n_1 }
 \Gamma^{\bar{c} \bar{c} cc}_{\Lambda} ( \sigma_2^{\prime} \omega_2^{\prime} , 
  \sigma_1^{\prime}  \omega_1^{\prime} ;  \sigma_2 \omega_2 ,  \sigma_1 \omega_1 ) + {\cal{O}} ( 1/N ),
 \end{align}
where
 \begin{align}
 & \Gamma^{\bar{c} \bar{c} cc}_{\Lambda} ( \sigma_1^{\prime} \omega_1^{\prime} , 
  \sigma_2^{\prime} \omega_2^{\prime} ;
  \sigma_2 \omega_2 ,  \sigma_1 \omega_1 ) =
 \nonumber
 \\
&  \lim_{ N \rightarrow \infty}
 \Gamma^{\bar{c} \bar{c} cc}_{\Lambda} ( n_1 \sigma_1^{\prime} \omega_1^{\prime} , 
 n_2 \sigma_2^{\prime} \omega_2^{\prime} ;
 n_2 \sigma_2 \omega_2 , n_1 \sigma_1 \omega_1 ) .
 \label{eq:gamma4fermilarge}
 \end{align}
Note that to leading order in $1/N$
the fermionic four-point vertex gives the following contribution to the vertex expansion of the average effective action
defined in Eq.~\eqref{eq:vertexexp},
\begin{widetext}
\begin{align}
 & \frac{1}{  (2!)^2    N \beta^4} \sum_{ n_1^{\prime} n_2^{\prime} n_2 n_1 }  
 \sum_{\sigma_1^{\prime} \sigma_2^\prime \sigma_2 \sigma_1} 
 \sum_{ \omega_1^{\prime} 
 \omega_2^{\prime} \omega_2 \omega_1 } \beta \delta_{ \omega_1^{\prime} + \omega_2^{\prime} ,
  \omega_2 + \omega_1 } 
 \Gamma_{\Lambda}^{\bar{c} \bar{c} cc} ( 1^{\prime},
 2^{\prime}  ; 2  , 1 )
\bar{c}_{ n_1^{\prime} \sigma_1^{\prime} \omega_1^{\prime}} \bar{c}_{ n_2^{\prime} 
 \sigma_2^{\prime} \omega_2^{\prime}}
 c_{ n_2 \sigma_2 \omega_2 } c_{ n_1 \sigma_1\omega_1 } 
 \nonumber
 \\
 \approx {} &
  \frac{ N}{ 2! \beta^4}
  \sum_{\sigma_1^{\prime} \sigma_2^\prime \sigma_2 \sigma_1} 
 \sum_{ \omega_1^{\prime} 
 \omega_2^{\prime} \omega_2 \omega_1 } \beta \delta_{ \omega_1^{\prime} + \omega_2^{\prime} ,
 \omega_2 + \omega_1 } \Gamma^{ \bar{c} \bar{c} cc}_{\Lambda} ( 
 \sigma_1^{\prime} \omega_1^{\prime},  \sigma_2^{\prime} \omega_2^{\prime} ; \sigma_2 \omega_2 , \sigma_1 \omega_1 ) \Psi_{ \sigma_1^{\prime} \omega_1^{\prime} \sigma_1 \omega_1 } \Psi_{\sigma_2^{\prime} \omega_2^{\prime} \sigma_2 \omega_2} ,
 \end{align}
\end{widetext}
where we have introduced the site-averaged composite field
 \begin{equation}
\Psi_{ \sigma^{\prime} \omega^{\prime} \sigma \omega } = \frac{1}{N} \sum_n \bar{c}_{ n \sigma^{\prime} \omega^{\prime}} c_{ n \sigma \omega }.
 \label{eq:Psidef}
 \end{equation}
This field  resembles the collective Hubbard-Stratonovich field introduced
in the path integral derivation of the Dyson-Schwinger equations from the large-$N$ 
saddle point \cite{Esterlis19,Sachdev15,Kitaev17}. However, in Eq.~(\ref{eq:Psidef}) the symbols 
 $\bar{c}_{ n \sigma^{\prime} \omega^{\prime}}$ and $ c_{ n \sigma \omega }$ represent the
source-dependent expectation values of the corresponding Grassmann fields,
so that the above $\Psi_{ \sigma^{\prime} \omega^{\prime} \sigma \omega }$
cannot be identified with the Hubbard-Stratonovich field introduced
in the path integral approach.

To solve our flow equations for the two-point vertices for large $N$ we also need the leading large-$N$ limit of the mixed four-point vertex,
 \begin{align}
 & \Gamma_{\Lambda}^{ \bar{c} c \phi \phi} ( n^{\prime} \sigma \omega^{\prime} ; n \sigma \omega; 
 k_1 \Omega_1 ,  k_2 \Omega_2 ) =
 \nonumber
 \\
 & \delta_{ n^{\prime} n} \delta_{k_1 k_2 } 
 \Gamma_{\Lambda}^{ \bar{c} c \phi \phi} ( \sigma \omega^{\prime} ; \sigma \omega; \Omega_1 , \Omega_2 )
 + {\cal{O}} ( 1 /N),
 \end{align}
where
 \begin{align}
 & \Gamma_{\Lambda}^{ \bar{c} c \phi \phi} ( \sigma \omega^{\prime} ; \sigma \omega; \Omega_1 , \Omega_2 ) =
 \nonumber
 \\
 & \lim_{N \rightarrow \infty} 
\Gamma_{\Lambda}^{ \bar{c} c \phi \phi} ( n \sigma \omega^{\prime} ; n \sigma \omega; 
 k \Omega_1 ,  k \Omega_2 ) .
 \label{eq:gamma4mixedlarge}
 \end{align}
Finally, the purely bosonic  four-point vertex has for large $N$ the form
 \begin{align}
 & \Gamma_{\Lambda}^{ \phi \phi \phi \phi} ( k_1  \Omega_1 , k_2 \Omega_2 , 
 k_3 \Omega_3 ,  k_4 \Omega_4 ) =
 \nonumber
 \\
 & \delta_{ k_1 k_2 } \delta_{k_3 k_4 } 
 \Gamma_{\Lambda}^{ \phi \phi \phi \phi} ( \Omega_1 , \Omega_2,  \Omega_3 , \Omega_4 )
 \nonumber
 \\
 + & \delta_{ k_1 k_3} \delta_{k_2 k_4 } 
 \Gamma_{\Lambda}^{ \phi \phi \phi \phi} ( \Omega_1 , \Omega_3,  \Omega_2 , \Omega_4 )
 \nonumber
 \\
 + & \delta_{ k_1 k_4} \delta_{k_2 k_3 } 
 \Gamma_{\Lambda}^{ \phi \phi \phi \phi} ( \Omega_1 , \Omega_4,  \Omega_2 , \Omega_3 )
 + {\cal{O}} ( 1 /N),
 \end{align}
with
 \begin{align}
 & \Gamma_{\Lambda}^{ \phi \phi \phi \phi} ( \Omega_1 , \Omega_2,  \Omega_3 , \Omega_4 ) =
 \nonumber
 \\
 & \lim_{N \rightarrow \infty} 
\Gamma_{\Lambda}^{ \phi \phi \phi \phi} ( k_1  \Omega_1 , k_1 \Omega_2 , 
 k_2 \Omega_3 ,  k_2 \Omega_4 ) .
 \end{align}
It turns out that the bosonic four-point vertex does not contribute to the flow of the
two-point vertices to leading order in $1/N$ because  this vertex vanishes in the bare action
and is not generated  by the FRG flow to this order.

\begin{figure}[tb]
 \begin{center}
  \centering
\vspace{7mm}
 \includegraphics[width=0.45\textwidth]{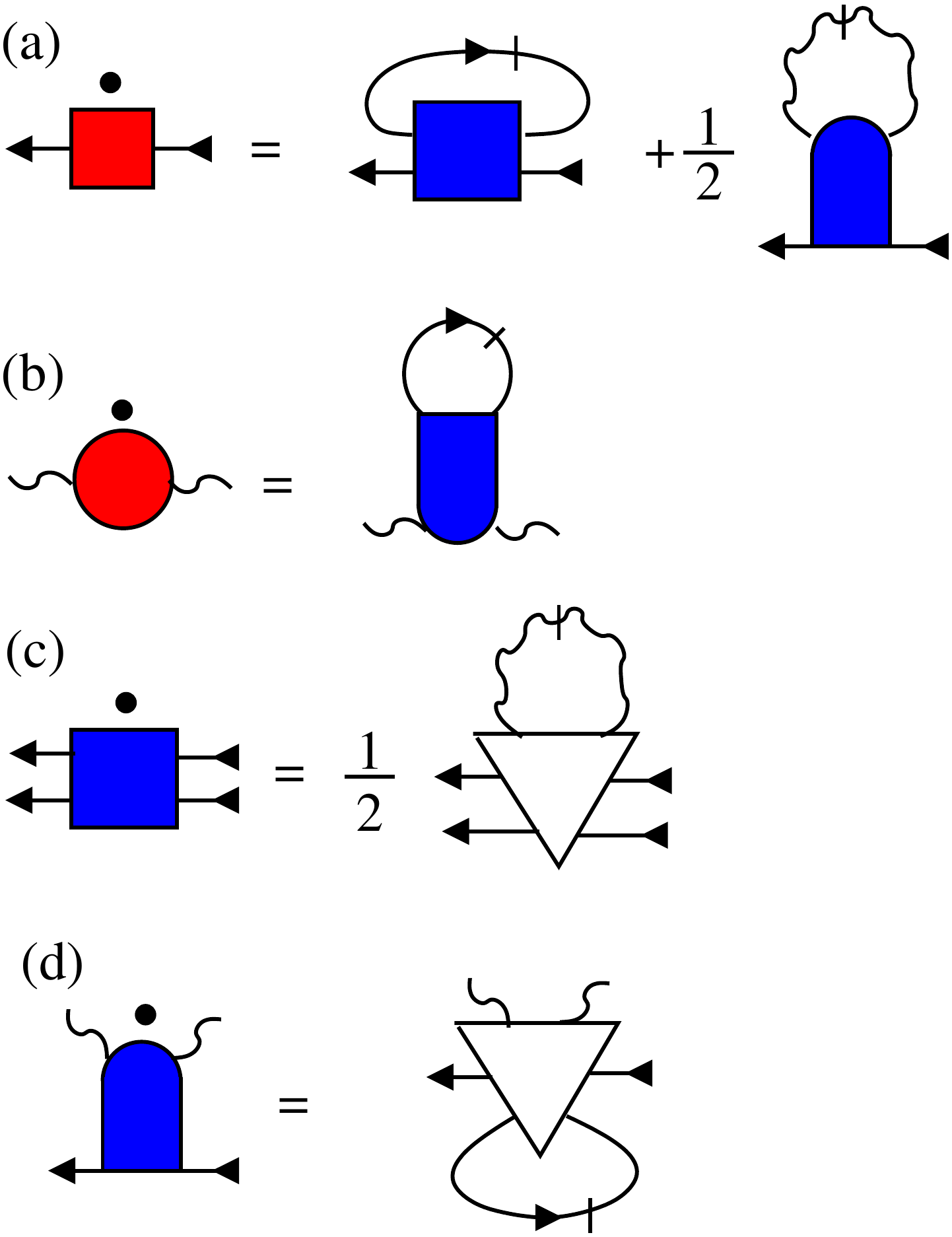}
   \end{center}
  \caption{ 
Graphical representation of our large-$N$ truncation of the hierarchy of
FRG flow equations for the YSYK model. 
(a) fermionic self-energy, Eq.~\eqref{eq:flowselfaverage}; 
(b) bosonic self-energy, Eq.~\eqref{eq:flowPiaverage}; 
(c) fermionic four-point vertex, Eq.~\eqref{eq:truncflowfour1}; 
(d) mixed four-point vertex, Eq.~\eqref{eq:truncflowfour2}. 
The empty triangle represents the large-$N$ limit of the 
symmetrized bare six-point vertex given in Eq.~\eqref{eq:sixpointinit}.
}
\label{fig:flow}
\end{figure}

It is now easy to see that the loop contributions $L^{\rm pp}_{\Lambda}$, $L^{\rm ph}_{\Lambda}$ and $L^{\rm bos}_{\Lambda} $ defined in Eqs.~(\ref{eq:Lpp} -- \ref{eq:Lbos}) do not contribute to the flow of the fermionic four-point vertex to leading order in $1/N$ \cite{Smit22}.  
Consider first the particle-particle loop 
$L^{\rm pp}_{\Lambda} ( 1^{\prime} , 2^{\prime} ; 2 , 1 )$ in Eq.~(\ref{eq:Lpp}) 
for the relevant index combinations $ n_1^{\prime} = n_1, n_2^{\prime} = n_2$ or $n_1^{\prime} = n_2, n_2^{\prime} = n_1$. Keeping in mind that for large $N$ the propagators are diagonal in the flavor indices, we see that the flavor sums in  Eq.~\eqref{eq:Lpp} collapses to $ 2 = {\cal{O}} (1) $ terms,
so that $L^{\rm pp}_{\Lambda} ( 1^{\prime} , 2^{\prime} ; 2 , 1 ) = {\cal{O}} ( 1/ N^2)$. 
Moreover, in  $L^{\rm ph}_{\Lambda} ( 1^{\prime} , 2^{\prime} ; 2 , 1 )$
and  $L^{\rm bos}_{\Lambda} ( 1^{\prime} , 2^{\prime} ; 2 , 1 )$ only one of the flavor sums collapses, so that these contributions are of order $1/N$. By the same argument, 
the leading large-$N$ contribution in the FRG flow equations for any $n$-point vertex 
is given by the term where two legs of the $(n+2)$-point vertex are joined by a single 
propagator line. In the flow equations for the four-point vertices, we therefore have to consider only the contributions from the six-point vertices.
Given the fact that the 
 bare action $S_6$ given in Eq.~\eqref{eq:S6freq} depends only on the  mixed 
six-point vertex $\Gamma_{\Lambda}^{ \bar{c} \bar{c} cc \phi \phi} $
with initial value given in Eq.~\eqref{eq:sixpointsym},
to leading order in $1/N$ 
we can neglect the FRG flow of 
all other six-point vertices. 
If we insert this initial value on the right-hand side of the flow equation
\eqref{eq:flow4point}, we see that only the site-average
of the six-point vertex given in Eq.~\eqref{eq:sixpointinit}
contributes. Moreover, following the  analysis 
of the diagrams in the flow equation for the fermionic four-point vertex
presented in the paragraph after Eq.~\eqref{eq:flow4point},  and using 
the
fact that the bare action does not contain any eight-point vertices, we see that to leading order
in $1/N$ the mixed six-point vertex is not renormalized. 
We thus arrive at the following large-$N$ truncation
of the formally exact FRG flow equations for the YSYK model:
The flow of the site-diagonal two-point vertices is related to the
large-$N$ limits of four-point vertices
defined in Eqs.~\eqref{eq:gamma4fermilarge} and
\eqref{eq:gamma4mixedlarge} as follows,
\begin{align}
 \partial_{\Lambda} \Sigma_{\Lambda} (  \omega ) = {} & 
 \frac{1}{\beta}  
 \sum_{ \sigma^{\prime} \omega^{\prime} } \dot{G}_{\Lambda} (  
 \omega^{\prime} ) 
 \Gamma^{\bar{c} \bar{c} cc}_{\Lambda} (   \sigma  \omega ,  \sigma^{\prime} \omega^{\prime}  ; 
 \sigma^{\prime} \omega^{\prime} ,  \sigma  \omega )
 \nonumber
 \\
 & +  \frac{1}{2  \beta} 
 \sum_{ \Omega } \dot{F}_{\Lambda} (  \Omega ) 
 \Gamma^{\bar{c} c \phi \phi }_{\Lambda} (  \sigma  \omega ;  \sigma \omega ;   -\Omega , \Omega ),
 \label{eq:flowselfaverage}
  \\
 \partial_{\Lambda} \Pi_{\Lambda} (  \Omega ) = {} & 
 \frac{N}{M \beta} 
 \sum_{ \sigma \omega } \dot{G}_{\Lambda} (  \omega ) 
\Gamma^{\bar{c} {c} \phi \phi}_{\Lambda} (   \sigma  \omega ; \sigma \omega  ; - \Omega,  \Omega ),
 \nonumber
 \\
 \label{eq:flowPiaverage}
 \end{align}
where the single-scale propagators are
 \begin{align}
 \dot{G}_{\Lambda} ( \omega ) & =  - G^2_{\Lambda} ( \omega ) \partial_{\Lambda} G_{0 , \Lambda}^{-1} ( \omega ) =  G^2_{\Lambda} ( \omega ) \partial_{\Lambda}
R_{\Lambda} ( \omega ),
\phantom{a}
 \label{eq:singleG}
 \\
 \dot{F}_{\Lambda} ( \Omega ) & = -   
 F^2_{\Lambda} ( \Omega ) \partial_{\Lambda} F_{0 , \Lambda}^{-1} ( \Omega ) =  
 - F^2_{\Lambda} ( \Omega ) \partial_{\Lambda}
 R^{\phi}_{\Lambda} ( \Omega ),
 \label{eq:singleF}
 \end{align}
and the scale-dependent propagators are related to the corresponding self-energies via the regularized Dyson equations 
 \begin{align}
 G_{\Lambda} ( \omega ) & =  \frac{1}{ G_{ 0 , \Lambda}^{-1} ( \omega ) - \Sigma_{\Lambda} ( \omega ) } 
 \nonumber
 \\ 
 & =
 \frac{1}{ i \omega + \mu 
- R_{\Lambda} ( \omega )  - \Sigma_{\Lambda} ( \omega )     },
 \\
 F_{\Lambda} ( \Omega ) & = \frac{1}{ F_{0, \Lambda}^{-1} ( \Omega ) + \Pi_{\Lambda} ( \Omega ) }
 \nonumber
 \\
 & =  \frac{1}{ | \Omega | + \Delta + R^{\phi}_{\Lambda} ( \Omega )  + \Pi_{\Lambda} ( \Omega ) }.
 \end{align}
The four-point vertices on the right-hand side of the flow equations
\eqref{eq:flowselfaverage} and \eqref{eq:flowPiaverage} for the 
two-point vertices satisfy 
 \begin{align}
 &
 \partial_{\Lambda}  \Gamma^{\bar{c} \bar{c} cc}_{\Lambda} (   \sigma_1^{\prime}  \omega_1^{\prime}
 ,  \sigma_2^{\prime} \omega_2^{\prime}  ; 
 \sigma_2 \omega_2 ,  \sigma_1  \omega_1 ) 
 = \frac{1}{2 \beta} \sum_{\Omega} \dot{F}_{\Lambda} ( \Omega ) 
 \nonumber
 \\
 & \times 
\Gamma_0^{\bar{c} \bar{c} cc \phi \phi} (   \sigma_1^{\prime}  \omega_1^{\prime}
 ,  \sigma_2^{\prime} \omega_2^{\prime}  ; 
 \sigma_2 \omega_2 ,  \sigma_1  \omega_1 ; - \Omega , \Omega),
 \label{eq:truncflowfour1}
 \end{align}
and
 \begin{align}
 &
 \partial_{\Lambda}  \Gamma^{\bar{c}  c \phi \phi}_{\Lambda} (   \sigma^{\prime}  
\omega^{\prime}
 ;  \sigma \omega  ;  \Omega_1 ,  \Omega_2 ) =
 \frac{1}{ \beta} \sum_{ \sigma_1 \omega_1} \dot{G}_{\Lambda} (  \omega_1 ) 
 \nonumber
 \\
 & \times
\Gamma_0^{\bar{c} \bar{c} cc \phi \phi} (   \sigma^{\prime}  \omega^{\prime}
 ,  \sigma_1 \omega_1  ; 
 \sigma_1 \omega_1 ,  \sigma  \omega ; \Omega_1 , \Omega_2 ),
 \label{eq:truncflowfour2}
 \end{align}
where
the large-$N$ limit of the initial value of the site-averaged mixed six-point vertex
$\Gamma_0^{\bar{c} \bar{c} cc \phi \phi} (   \ldots )$
is given in Eq.~\eqref{eq:sixpointinit}.
A graphical representation of the above large-$N$ truncation of the
FRG flow equations is shown in
Fig.~\ref{fig:flow}.
Substituting 
our explicit expression \eqref{eq:sixpointinit} for the initial value of the
site averaged six-point vertex 
into Eqs.~\eqref{eq:truncflowfour1} and \eqref{eq:truncflowfour2},
we see that the
fermionic four-point vertex 
in the flow equation~\eqref{eq:flowselfaverage} is spin-diagonal
while the mixed four-point vertex in Eq.~\eqref{eq:flowPiaverage} is actually
independent of the spin label $\sigma$ so that we can write
 \begin{align}
 \Gamma^{\bar{c} \bar{c} cc}_{\Lambda} (   \sigma  \omega ,  \sigma^{\prime} \omega^{\prime}  ; 
 \sigma^{\prime} \omega^{\prime} ,  \sigma  \omega )
 & = \delta_{\sigma \sigma^{\prime} } 
 \Gamma^{\bar{c} \bar{c} cc}_{\Lambda} (   \omega ,  \omega^{\prime}  ; 
 \omega^{\prime} ,   \omega ),
 \\
 \Gamma^{\bar{c} c \phi \phi }_{\Lambda} (  \sigma  \omega ;  \sigma \omega ;   -\Omega , \Omega ) 
 & =
 \Gamma^{\bar{c} c \phi \phi }_{\Lambda} (   \omega ;   \omega ;   -\Omega , \Omega ).
 \end{align}
With this notation the flow equations (\ref{eq:flowselfaverage}, \ref{eq:flowPiaverage}) for the self-energies can be written as
 \begin{align}
 \partial_{\Lambda} \Sigma_{\Lambda} (  \omega ) = {} & 
 \frac{1}{\beta}  
 \sum_{  \omega^{\prime} } \dot{G}_{\Lambda} (  
 \omega^{\prime} ) 
 \Gamma^{\bar{c} \bar{c} cc}_{\Lambda} (   \omega ,   \omega^{\prime}  ; 
  \omega^{\prime} ,    \omega )
 \nonumber
 \\
 & +   \frac{1}{2  \beta} 
 \sum_{ \Omega } \dot{F}_{\Lambda} (  \Omega ) 
 \Gamma^{\bar{c} c \phi \phi }_{\Lambda} (    \omega ;  \omega ;   -\Omega , \Omega ),
 \label{eq:flowselfaverage2}
  \\
 \partial_{\Lambda} \Pi_{\Lambda} (  \Omega ) = {} & 
 \frac{p}{ \beta}
 \sum_{  \omega } \dot{G}_{\Lambda} (  \omega ) 
\Gamma^{\bar{c} {c} \phi \phi}_{\Lambda} (     \omega ; \omega  ; - \Omega,  \Omega ),
 \label{eq:flowPiaverage2}
 \end{align}
while the flow of the scale-dependent  four-point vertices is
 \begin{align}
   \partial_{\Lambda}  \Gamma^{\bar{c} \bar{c} cc}_{\Lambda} (     \omega
  ,  \omega^{\prime}  ;  \omega^{\prime} ,    \omega ) 
  &  =   g^2    \dot{F}_{\Lambda} ( \omega - \omega^{\prime} ),
 \label{eq:flow4a}
  \\
  \partial_{\Lambda}  \Gamma^{\bar{c}  c \phi \phi}_{\Lambda} ( 
\omega ;   \omega  ;  - \Omega ,  \Omega ) 
  &   =  
   g^2  
 \bigl[  \dot{G}_{\Lambda} ( \omega + \Omega )  
+   \dot{G}_{\Lambda} ( \omega - \Omega )
 \bigr]. 
 \label{eq:flow4b}
 \end{align}
Here 
\begin{equation}
 g^2 = g_1^2 + g_2^2
 \end{equation}
is the relevant bare coupling and
we have introduced the parameter
 \begin{equation}
 p = \frac{ ( 2 S+1 ) N}{M}  = \frac{2 N}{M},
 \label{eq:pdef}
 \end{equation}
where $2S+1 =2$ is the spin-degeneracy for spin $S=1/2$.

The four coupled flow equations~(\ref{eq:flowselfaverage2}--\ref{eq:flow4b}) 
uniquely determine  the scale-dependent 
fermionic and bosonic self-energies and  
the two relevant interaction vertices. 
We emphasize that Eqs.~(\ref{eq:flowselfaverage2}--\ref{eq:flow4b})
have been obtained from the formally exact hierarchy of flow equations for the irreducible vertices
of the YSYK model by retaining all terms which have  a  
finite limit for $N \rightarrow \infty$ and $M \rightarrow \infty$. 

We now show that our FRG flow equations ~(\ref{eq:flowselfaverage2}--\ref{eq:flow4b}) 
 can be reduced to the usual DS equations for
the self-energies of the YSYK model if we
use the so-called Katanin substitution \cite{Katanin04} to
replace the single-scale propagators on the right-hand sides by the
total scale-derivatives of the propagators, 
 \begin{align}
  \dot{G}_{\Lambda} ( \omega ) & \rightarrow \partial_\Lambda G_{\Lambda} ( \omega )
  = \dot{G}_{\Lambda} ( \omega ) + G^2_{\Lambda} ( \omega ) \partial_{\Lambda} \Sigma_{\Lambda} ( \omega )
  \label{eq:KataninG}
  \\
  \dot{F}_{\Lambda} ( \Omega ) & \rightarrow \partial_\Lambda F_{\Lambda} ( \Omega ) =
    \dot{F}_{\Lambda} ( \Omega ) - F^2_{\Lambda} ( \Omega ) \partial_{\Lambda} \Pi_{\Lambda} ( \Omega ) .
    \label{eq:KataninF}
  \end{align}
With this substitution the right-hand sides of 
the flow equations (\ref{eq:flow4a}, \ref{eq:flow4b})
become total $\Lambda$-derivatives so that we may trivially integrate both sides over the flow parameter $\Lambda$. Taking into account that
for  $\Lambda_0 \rightarrow \infty$ the regularized 
propagators vanish we obtain
\begin{equation}
   \Gamma^{\bar{c} \bar{c} cc}_{\Lambda} (     \omega
 ,  \omega^{\prime}  ; 
  \omega^{\prime} ,    \omega ) 
  =    g^2   {F}_{\Lambda} ( \omega - \omega^{\prime} ),
 \label{eq:fourint1}
  \end{equation}
and
 \begin{equation}
  \Gamma^{\bar{c}  c \phi \phi}_{\Lambda} (  
\omega ;  \omega  ;  - \Omega ,  \Omega ) 
    =  
   g^2  
 \bigl[  {G}_{\Lambda} ( \omega + \Omega ) 
+   {G}_{\Lambda} ( \omega - \Omega )
 \bigr]. 
 \label{eq:fourint2}
 \end{equation}
Substituting these equations into the flow equations \eqref{eq:flowselfaverage} and \eqref{eq:flowPiaverage}
for the two-point functions we obtain
 \begin{align}
 \partial_{\Lambda} \Sigma_{\Lambda} ( \omega )
 & = \frac{ g^2}{\beta} \sum_{\Omega}
 [ \dot{F}_{\Lambda} ( \Omega ) G_{\Lambda} ( \omega - \Omega )
 \nonumber
 \\
 & \hspace{12mm}
 + {F}_{\Lambda} ( \Omega ) \dot{G}_{\Lambda} ( \omega - \Omega ) ],
 \label{eq:flowintSigma}
 \\
 \partial_{\Lambda} \Pi_{\Lambda} ( \Omega )
 & = p \frac{ g^2 }{\beta} \sum_{\omega}
 [ \dot{G}_{\Lambda} ( \omega ) G_{\Lambda} ( \omega - \Omega )
 \nonumber
 \\
 & \hspace{14mm}
 +  {G}_{\Lambda} ( \omega ) \dot{G}_{\Lambda} ( \omega - \Omega ) ].
 \label{eq:flowintPi}
 \end{align}
Using once more the Katanin substitution (\ref{eq:KataninG}, \ref{eq:KataninF}) to transform the right-hand side of the above equations
into total $\Lambda$-derivatives and integrating both sides over $\Lambda$ we obtain for $\Lambda
\rightarrow 0$ the 
DS equations for the YSYK model~\cite{Esterlis19,Classen21},
 \begin{align}
 \Sigma ( \omega )
 & = \frac{ g^2 }{\beta} \sum_{\Omega}
 {F}  ( \Omega ) G ( \omega - \Omega ),
 \label{eq:DSSigma}
 \\
 \Pi ( \Omega ) & = p \frac{ g^2}{\beta} \sum_{\omega}
 G ( \omega )  G ( \omega - \Omega ).
  \hspace{7mm}
 \label{eq:DSPi}
 \end{align} 
We conclude that our large-$N$ truncation of the FRG flow
equations in combination with the Katanin substitution 
is equivalent to the DS equations
obtained from the large-$N$ saddle point
of a suitably defined functional integral representation
of the YSYK model. 
Note that without the Katanin substitution
our large-$N$ FRG flow equations are {\it{not}} equivalent 
to the DS equations (\ref{eq:DSSigma}) and (\ref{eq:DSPi})
because the Katanin substitution in Eqs.~(\ref{eq:KataninG}, \ref{eq:KataninF})
re-sums higher orders in the coupling $g^2$.
However, for the SYK model we have shown in Ref.~[\onlinecite{Smit21}] that
the Katanin substitution does not modify the value of the  fermionic anomalous dimension.
Whether this is also true for the YSYK model is an interesting question beyond the scope of this work.

%
%

\section{Low-energy expansion}
\label{sec:flow}

Our aim is to determine
possible fixed points of the
system of FRG flow equations given in 
Eqs.~(\ref{eq:flowselfaverage2}--\ref{eq:flow4b})
at zero temperature, $ \beta^{ - 1 } = 0 $.
As usual, we anticipate that the fixed points are determined by the leading low-energy behavior of the vertices, so that
for our purpose it is sufficient to expand the
irreducible self-energies to linear order in the frequencies.
Taking the possibility of dissipative terms proportional to
$| \omega |$ and $ | \Omega |$ into account, the expansion is of the form
\begin{align}
 \Sigma_{\Lambda} ( \omega ) & = \Sigma_{\Lambda} ( 0 )  - A_{\Lambda} | \omega | 
 - ( B_{\Lambda} -1 ) i \omega +
  {\cal{O}} ( \omega^2 ),
  \label{eq:sigmaexpansion}
 \\
 \Pi_{\Lambda} ( \Omega ) & = \Pi_{\Lambda} (0 ) + ( Y_{\Lambda}^{-1} -1 ) 
 | \Omega | + {\cal{O}} ( \Omega^2 ),
  \label{eq:piexpansion}
 \end{align}
where $A_{\Lambda}$, $B_{\Lambda}$, and $Y_{\Lambda}$ are dimensionless.
The coupling $A_{\Lambda}$ parametrizes the asymmetry in the
fermion spectral function which is expected to emerge for finite values of the chemical
potential $\mu$ \cite{Wang20a}. 
The  usual fermionic wave-function renormalization factor $Z_{\Lambda}$ is then given by 
 \begin{equation}
 Z_{\Lambda} = \frac{1}{ \sqrt{ A_{\Lambda}^2 + B_{\Lambda}^2 } },
 \end{equation}
which defines the scale-dependent fermionic anomalous dimension  via
 \begin{equation}
 \eta_{\Lambda} = \frac{ \Lambda \partial_{\Lambda} Z_{\Lambda}}{Z_\Lambda}.
 \end{equation} 
Using for simplicity a sharp frequency regulator,
the low-energy form of the scale-dependent fermionic propagator and the
corresponding single-scale propagator can then be written as \cite{MorrisLemma}
 \begin{subequations}
 \begin{align}
 G_{\Lambda} ( \omega ) &  = \frac{ Z_{\Lambda} \Theta ( | \omega | - \Lambda )}{
   a_{\Lambda} | \omega|  +  i b_{\Lambda}  \omega  + \mu_{\Lambda}},
 \label{eq:Gsharp}  
 \\
 \dot{G}_{\Lambda} ( \omega ) &  = - \frac{ Z_{\Lambda} \delta ( | \omega | - \Lambda )}{
  a_{\Lambda} \Lambda +   i   b_\Lambda \Lambda  {\rm sgn } {\omega}   +  \mu_{\Lambda}},
 \label{eq:dotGsharp}
 \end{align}
 \end{subequations}
where
 \begin{subequations}
 \begin{align} 
 a_\Lambda & =  Z_{\Lambda} A_{\Lambda} = \frac{ A_{\Lambda}}{   \sqrt{ A_{\Lambda}^2 + B_{\Lambda}^2 }},
 \\
 b_\Lambda & =  Z_{\Lambda} B_{\Lambda} = \frac{ B_{\Lambda}}{   \sqrt{ A_{\Lambda}^2 + B_{\Lambda}^2 }},
 \end{align}
 \end{subequations}
and
 \begin{equation}
 \mu_{\Lambda}  =  Z_{\Lambda} [ \mu - \Sigma_{\Lambda} (0) ].
 \end{equation}
Note that by construction $a_\Lambda^2 + b_\Lambda^2=1$.
Similarly, using also a sharp frequency cutoff for the bosons, the bosonic
propagator and the corresponding single-scale propagator are at low energies given by
\begin{subequations}
 \label{eq:FFsharp}
 \begin{align}
 F_{\Lambda} ( \Omega ) &   = \frac{ Y_{\Lambda} \Theta ( | \Omega | - \Lambda )}{
  | \Omega | + r_{\Lambda}},
 \\
 \dot{F}_{\Lambda} ( \Omega ) & =   - \frac{ Y_{\Lambda} \delta ( | \Omega | - \Lambda )}{ \Lambda  + r_{\Lambda}},
 \label{eq:dotFsharp}
 \end{align}
 \end{subequations}
where
 \begin{equation}
 r_{\Lambda} = Y_{\Lambda} [ \Delta + \Pi_{\Lambda} (0) ].
 \end{equation}
The logarithmic scale-dependence of the 
bosonic wave-function renormalization factor $Y_{\Lambda}$
defines the scale-dependent bosonic anomalous dimension
\begin{equation}
 \gamma_{\Lambda} = \frac{ \Lambda \partial_{\Lambda} Y_{\Lambda}}{Y_\Lambda}.
 \end{equation} 
For later convenience we introduce the dimensionless coupling
 \begin{equation}
 {u}_l = \frac{ Z_l^2 Y_l g^2}{\pi \Lambda^2},
 \label{eq:uldef}
 \end{equation}
which by definition satisfies the flow equation
 \begin{equation}
 \partial_l {u}_l = ( 2 - 2 \eta_l - \gamma_l ) {u}_l,
 \label{eq:uflow}
 \end{equation}
where $l = \ln ( \Lambda_0 / \Lambda )$ is the logarithmic flow parameter
and $\partial_l = - \Lambda \partial_{\Lambda}$.
From now on all quantities will be considered to be functions
of $l$ instead of $\Lambda$; in the case of dimensionless quantities we simply rename 
$\eta_{\Lambda } \rightarrow \eta_l$, $\gamma_{\Lambda} \rightarrow \gamma_l$,
$a_{\Lambda} \rightarrow a_l$ and $b_{\Lambda} \rightarrow b_l$, while the dimensionful couplings
$\mu_{\Lambda}$ and $r_{\Lambda}$ are
divided by $\Lambda$ to obtain the corresponding dimensionless rescaled couplings,
 \begin{align}
 \mu_l & = \frac{ \mu_{\Lambda}}{\Lambda} = \frac{ Z_{\Lambda} [ \mu - \Sigma_{\Lambda} (0) ]}{
 \Lambda},
 \\
 r_l & = \frac{ r_{\Lambda}}{\Lambda} = \frac{ Y_{\Lambda} [ \Delta + \Pi_{\Lambda} (0) ]}{
 \Lambda}.
 \end{align}
To find the fixed points of the renormalization group,
we introduce the dimensionless rescaled self-energies
 \begin{align}
 \tilde{\Sigma}_l ( \tilde{\omega} ) & = 
 \frac{ Z_{\Lambda} \Sigma_{\Lambda} ( \Lambda \tilde{\omega} ) }{ \Lambda},
 \\
 \tilde{\Pi}_l ( \tilde{\Omega} ) & = \frac{ Y_{\Lambda} \Pi_\Lambda ( \Lambda
 \tilde{\Omega} ) }{\Lambda} ,
 \end{align}
which depend on the dimensionless frequencies 
$\tilde{\omega} = \omega / \Lambda$ and $\tilde{\Omega} = \Omega / \Lambda$.
Using the  flow equation \eqref{eq:flowselfaverage2} and \eqref{eq:flowPiaverage2} for the
dimensionful self-energies we find that the corresponding dimensionless rescaled quantities satisfy
  \begin{align}
 \partial_l \tilde{\Sigma}_l ( \tilde{\omega} ) & =
 ( 1 - \eta_l - \tilde{\omega} \partial_{\tilde{\omega} } ) \tilde{\Sigma}_l ( \tilde{\omega} ) + \dot{\Sigma}_l ( \tilde{\omega} ),
  \label{eq:Sigmaflowscale}
 \\
\partial_l \tilde{\Pi}_l ( \tilde{\Omega} ) & =
 ( 1 - \gamma_l - \tilde{\Omega} \partial_{\tilde{\Omega} } ) \tilde{\Pi}_l ( \tilde{\Omega} ) + \dot{\Pi}_l ( \tilde{\Omega} ),
 \label{eq:Piflowscale}
 \end{align}
where
 \begin{align}
 \dot{\Sigma}_l ( \tilde{\omega} ) = {} & - Z_{\Lambda} \partial_{\Lambda} \Sigma_{\Lambda} ( \omega ) 
 \nonumber
 \\
 = {} & \int \frac{ d \tilde{\omega}^{\prime}}{ 2 \pi }
 \dot{\tilde{G}}_l ( \tilde{\omega}^{\prime} ) \tilde{\Gamma}^{ \bar{c} \bar{c} cc}_l (
 \tilde{\omega} , \tilde{\omega}^{\prime} ; \tilde{\omega}^{\prime} , \tilde{\omega} )
 \nonumber
 \\
 & + \frac{1}{2} \int \frac{ d \tilde{\Omega} }{ 2 \pi }
 \dot{\tilde{F}}_l ( \tilde{\Omega} ) \tilde{\Gamma}^{ \bar{c} c \phi \phi}_l (
 \tilde{\omega} , \tilde{\omega} ; - \tilde{\Omega} , \tilde{\Omega} ),
 \label{eq:Aldef}
  \\
 \dot{\Pi}_l ( \tilde{\Omega} ) = {} & - Y_{\Lambda} \partial_{\Lambda} \Pi_{\Lambda} ( \Omega )
 \nonumber
 \\
 = {} & p  \int \frac{ d \tilde{\omega} }{ 2 \pi }
 \dot{\tilde{G}}_l ( \tilde{\omega} ) \tilde{\Gamma}^{ \bar{c} c \phi \phi}_l (
 \tilde{\omega} , \tilde{\omega} ; - \tilde{\Omega} , \tilde{\Omega} ).
 \label{eq:Bldef}
 \end{align}
Here we have introduced the dimensionless rescaled four-point vertices,
  \begin{align}    
 \tilde{\Gamma}_l^{ \bar{c} \bar{c} cc } (  \tilde{\omega} , 
  \tilde{\omega}^{\prime} ;
  \tilde{\omega}^{\prime} , \tilde{\omega}) 
&  = \frac{ Z_{\Lambda}^2}{\Lambda}
 {\Gamma}_\Lambda^{ \bar{c} \bar{c} cc } (   \Lambda \tilde{\omega}  ,  \Lambda \tilde{\omega}^{\prime} ;
  \Lambda \tilde{\omega}^{\prime} ,  \Lambda \tilde{\omega}),
 \\
   \tilde{\Gamma}_l^{ \bar{c}{c} \phi \phi  } (  \tilde{\omega} ,  \tilde{\omega} ;
  - \tilde{\Omega} , \tilde{\Omega}) 
&   = \frac{ Z_{\Lambda} Y_{\Lambda}}{\Lambda}
 {\Gamma}_\Lambda^{ \bar{c}{c} \phi \phi  } (   \Lambda \tilde{\omega}  ;  \Lambda \tilde{\omega} ;
 - \Lambda \tilde{\Omega} , \Lambda \tilde{\Omega}),
 \end{align}
and the dimensionless single-scale propagators,
 \begin{align}
  \dot{\tilde{G}}_l ( \tilde{\omega} )  & =    - \frac{\Lambda^2}{ Z_{\Lambda}} \dot{G}_{\Lambda} ( \Lambda \omega )  \approx 
 \frac{\delta ( | \tilde{\omega} | -1 )}{ a_l + i b_l {\rm sgn} \omega      + \mu_l },
 \hspace{7mm}
 \\
  \dot{\tilde{F}}_l ( \tilde{\omega} )  & =    - \frac{\Lambda^2}{ Y_{\Lambda}} \dot{F}_{\Lambda} 
 ( \Lambda \omega )  \approx   \frac{\delta ( | \tilde{\Omega} | -1 )}{1  + {r}_l }.
 \end{align}
From the flow equations \eqref{eq:flow4a} and \eqref{eq:flow4b} we obtain for the flow of the rescaled four-point vertices
\begin{widetext}
 \begin{align}
 \partial_l \tilde{\Gamma}_l^{ \bar{c} \bar{c} cc } ( \tilde{\omega}_1^{\prime} ,  \tilde{\omega}_2^{\prime} ;
 \tilde{\omega}_2 ,  \tilde{\omega}_1 )  = {} &  
 ( 1 - 2 \eta_l   - \tilde{\omega}_{1}^{\prime} \partial_{\tilde{\omega}_1^{\prime}} 
 -  \tilde{\omega}_{2}^{\prime} \partial_{\tilde{\omega}_2^{\prime}} 
  - \tilde{\omega}_{2} \partial_{\tilde{\omega}_2}    -  \tilde{\omega}_{1} \partial_{\tilde{\omega}_1}  
) \tilde{\Gamma}_l^{ \bar{c} \bar{c} cc } ( \tilde{\omega}_1^{\prime} ,  \tilde{\omega}_2^{\prime} ;
 \tilde{\omega}_2 ,  \tilde{\omega}_1 )
\nonumber
 \\
 & + \pi \frac{ {u}_l}{2} \left[ \dot{\tilde{F}}_l ( \tilde{\omega}_1 - \tilde{\omega}_2^{\prime} )
 +   \dot{\tilde{F}}_l ( \tilde{\omega}_2 - \tilde{\omega}_1^{\prime} ) \right],
 \label{eq:Gamma4flow1}
 \\
 \partial_l \tilde{\Gamma}_l^{ \bar{c}c \phi \phi } (  \tilde{\omega}_1 ,  \tilde{\omega}_2 ;
  \tilde{\Omega}_1 ,  \tilde{\Omega}_2)  = {} &  
 ( 1 -  \eta_l - \gamma_l 
 - \tilde{\omega}_{1} \partial_{\tilde{\omega}_1} 
 -  \tilde{\omega}_{2} \partial_{\tilde{\omega}_2} 
  - \tilde{\Omega}_{1} \partial_{\tilde{\Omega}_1}    -  \tilde{\Omega}_{2} \partial_{\tilde{\Omega}_2}  
)  \tilde{\Gamma}_l^{ \bar{c}c \phi \phi } (  \tilde{\omega}_1 ,  \tilde{\omega}_2 ;
  \tilde{\Omega}_1 ,  \tilde{\Omega}_2) 
 \nonumber
 \\
 & +  \pi \frac{ {u}_l}{2} \left[ \dot{\tilde{G}}_l ( \tilde{\omega}_1 + \tilde{\Omega}_1 )
 +   \dot{\tilde{G}}_l ( \tilde{\omega}_2 + \tilde{\Omega}_1 ) 
 + \dot{\tilde{G}}_l ( \tilde{\omega}_1 + \tilde{\Omega}_2 )
 +   \dot{\tilde{G}}_l ( \tilde{\omega}_2 + \tilde{\Omega}_2 ) 
\right].
 \label{eq:Gamma4flow2}
 \end{align}
These linear first-order partial differential equations 
can be solved analytically \cite{Kopietz10}. For the external frequencies needed in Eqs.~\eqref{eq:Aldef} and \eqref{eq:Bldef} we obtain
 \begin{align}
 \tilde{\Gamma}_l^{ \bar{c} \bar{c} cc } ( \tilde{\omega} ,  \tilde{\omega}^{\prime} ;
 \tilde{\omega}^{\prime} ,  \tilde{\omega} ) = {} & e^{ \int_0^l d \tau ( 1 - 2 \eta_{\tau} )}
  \tilde{\Gamma}_0^{ \bar{c} \bar{c} cc } ( e^{-l} \tilde{\omega} ,  e^{-l} \tilde{\omega}^{\prime} ;
 e^{-l} \tilde{\omega}^{\prime} ,  e^{-l} \tilde{\omega} )
 \nonumber
 \\
 & + \pi \int_0^{l} d t e^{ \int_{ l-t}^l d \tau ( 1 - 2 \eta_{\tau} ) } 
 {u}_{ l -t } \dot{\tilde{F}}_{l-t} ( e^{ - t} ( \tilde{\omega} - \tilde{\omega}^{\prime} ) ),
 \\
\tilde{\Gamma}_l^{ \bar{c} c \phi \phi } ( \tilde{\omega} ;  \tilde{\omega} ;
 - \tilde{\Omega} ,  \tilde{\Omega} ) = {} & e^{ \int_0^l d \tau ( 1 -  \eta_{\tau} - \gamma_{\tau})}
  \tilde{\Gamma}_0^{ \bar{c}  c \phi \phi } ( e^{-l} \tilde{\omega} ;  e^{-l} \tilde{\omega} ;  - e^{-l} \tilde{\Omega} ,  e^{-l} \tilde{\Omega} )
 \nonumber
 \\
 & + \pi \int_0^{l} d t e^{ \int_{ l-t}^l d \tau ( 1 -  \eta_{\tau} - \gamma_{\tau} ) } 
 {u}_{ l -t } \left[ \dot{\tilde{G}}_{l-t} ( e^{ - t} ( \tilde{\omega} - \tilde{\Omega} ) )
 + \dot{\tilde{G}}_{l-t} ( e^{ - t} ( \tilde{\omega} + \tilde{\Omega} ) ) \right].
 \end{align}
After substituting these expressions into Eqs.~\eqref{eq:Aldef} and \eqref{eq:Bldef} and performing the
frequency integrations we obtain
 \begin{align}
 \dot{\Sigma}_l ( \tilde{\omega} ) & = \frac{1}{2} \int_0^l dt {u}_{l-t} 
 \Biggl\{ \frac{e^{ \int_{l-t}^l d \tau ( 1 - 2 \eta_{\tau} ) }}{ 1 + {r}_{l-t}}
 \left[ \frac{\delta ( e^{-t} | 1 - \tilde{\omega} | -1 ) }{
 \tilde{\mu}_l + i b_l } +
\frac{\delta ( e^{-t} | 1 + \tilde{\omega}  | -1 ) }{
 \tilde{\mu}_l - i b_l } \right]
 \nonumber
 \\
 & \hspace{24mm} + \frac{e^{ \int_{l-t}^l d \tau ( 1 -  \eta_{\tau} - \gamma_{\tau}) }}{ 1 + 
 {r}_{l}}
 \left[ \frac{\delta ( e^{-t} | 1 - \tilde{\omega}  | -1 ) }{
 \tilde{\mu}_{l-t} -  i   b_{l-t} {\rm sgn} ( 1 - \tilde{\omega}  ) } +
\frac{\delta ( e^{-t} | 1 + \tilde{\omega}  | -1 ) }{
\tilde{\mu}_{l-t} + i b_{l-t} {\rm{sgn}} ( 1 + \tilde{\omega} )  } \right]
 \Biggr\},
 \\
 \dot{\Pi}_l ( \tilde{\Omega} ) & =  p   \int_0^l dt {u}_{l-t}  
 e^{ \int_{l-t}^l d \tau ( 1 -  \eta_{\tau} - \gamma_{\tau}) }
 {\rm Re} \left[ \frac{\delta ( e^{-t} | 1 - \tilde{\Omega}  | -1 ) }{
 ( \tilde{\mu}_l + i b_l ) ( \tilde{\mu}_{l-t} + i b_{l-t} {\rm sgn} ( 1 - \tilde{\Omega} ) )} +
\frac{\delta ( e^{-t} | 1 + \tilde{\Omega}  | -1 ) }{
 ( \tilde{\mu}_l + i b_l ) ( \tilde{\mu}_{l-t} + i b_{l-t} {\rm sgn} (  1 + \tilde{\Omega} )  )}
 \right],
  \end{align}
where we have introduced the shifted rescaled chemical potential,
 \begin{equation}
 \tilde{\mu}_l = \mu_l + a_l.
 \end{equation}
Assuming $ | \tilde{\omega} | < 1$ and
$| \tilde{\Omega} | < 1$ and using the fact that in this case the  $\delta$-functions 
can be written as
 \begin{equation}
 \delta (   e^{-t} | 1 \pm  \tilde{\omega}  | -1 ) = \delta (  t -  \ln ( 1 \pm \tilde{\omega} )),
 \end{equation}
we can now carry out the $t$-integrations. 
Defining $L_{\tilde{\omega}} = \ln ( 1 + | \tilde{\omega} | )$ and assuming $l > L_{\tilde{\omega} }$ and $l > L_{ \tilde{\Omega} }$ we finally obtain
 \begin{align}
 \dot{\Sigma}_l ( \tilde{\omega} ) & = \frac{ {u}_{ l - L_{ \tilde{\omega} }}}{2} 
 \left[    
\frac{e^{ \int_{l- L_{ \tilde{\omega} } }^l d \tau ( 1 - 2 \eta_{\tau} ) }}{ ( 1 + 
 {r}_{l- L_{  \tilde{\omega}}  })( \tilde{\mu}_l
 - i b_l {\rm sgn }  \omega ) }
+ \frac{e^{ \int_{l- L_{ \tilde{\omega} } }^l d \tau ( 1 -  \eta_{\tau} - \gamma_{\tau}) }}{ 
( 1 + {r}_{l} )( 
 \tilde{\mu}_{ l - L_{  \tilde{\omega} } }
 +  i  b_{l - L_{\tilde{\omega}}} {\rm sgn} \omega    ) }
 \right],
 \label{eq:sigmadotres}
 \\
 \dot{\Pi}_l ( \tilde{\Omega} ) & = p  {u}_{ l - L_{ \tilde{\Omega} } } 
 e^{ \int_{l- L_{ \tilde{\Omega} } }^l d \tau ( 1 -  \eta_{\tau} - \gamma_{\tau}) }
 {\rm Re} \left[ \frac{1}{ ( \tilde{\mu}_l + i b_l )( \tilde{\mu}_{ l - L_{  \tilde{\Omega} } } 
 + i b_{ l - L_{ \tilde{\Omega} } } ) }
 \right].
 \label{eq:pidotres}
 \end{align}
\end{widetext}

The flow equations for $\mu_l$ and ${r}_l$ can now be obtained 
from Eqs.~\eqref{eq:sigmadotres} and \eqref{eq:pidotres}
by setting the external frequencies equal to zero,
 \begin{align}
 \partial_l \mu_l & = ( 1 - \eta_l ) \mu_l - \dot{\Sigma}_l ( 0 )
 \nonumber 
 \\
 &  = ( 1 - \eta_l ) \mu_l - {u}_l \frac{ \tilde{\mu}_l }{ ( 1 + {r}_l ) (  
b_l^2 + \tilde{\mu}_l^2   ) },
 \label{eq:muflow}
 \\
 \partial_l {r}_l & = ( 1 - \gamma_l ) {r}_l + \dot{\Pi}_l (0)
\nonumber
 \\
 & =  
 ( 1 - \gamma_l ) {r}_l  - p {u}_l 
 \frac{ b_l^2 - \tilde{\mu}_l^2}{ ( b_l^2 + \tilde{\mu}_l^2 )^2 }.
 \label{eq:rflow}
 \end{align}
To determine the scale-dependent anomalous dimensions $\eta_l$ and $\gamma_l$ we need the  
linear terms in the expansions of $\dot{\Sigma}_l ( \tilde{\omega} )$ and $\dot{\Pi}_l ( \tilde{\Omega} )$,
 \begin{align}
 \dot{\Sigma}_l ( \tilde{\omega} ) & = \dot{\Sigma}_l ( 0 ) - \alpha_l | \tilde{\omega} | - 
 \beta_l  i \tilde{\omega} + {\cal{O}} ( \tilde{\omega}^2 ),
 \\
 \dot{\Pi}_l ( \tilde{\Omega} ) & = \dot{\Pi}_l ( 0 ) + \gamma_l | \tilde{\Omega} | + {\cal{O}} ( \tilde{\Omega}^2 ).
 \end{align} 
The fermionic anomalous dimension is then given by
 \begin{equation}
 \eta_l = a_l \alpha_l + b_l \beta_l,
 \end{equation}
where $a_l$ and $b_l$ satisfy 
 \begin{align}
  \partial_l a_l & = - \eta_l a_l + \alpha_l,
 \label{eq:aflow}
 \\
 \partial_l b_l & = - \eta_l b_l + \beta_l,
 \end{align}
with
  \begin{align}
 \alpha_l = {} & \frac{ {u}_l}{2 ( 1 + {r}_l )}
 \Biggl\{ \frac{ \tilde{\mu}_l}{ ( b_l^2 + \tilde{\mu}_l^2 )}
 \left[  2 - \eta_l - \gamma_l -  \frac{ \partial_l {r}_l }{ 1 +  {r}_l } \right]
 \nonumber
 \\
 & + \frac{ ( b_l^2 - \tilde{\mu}_l^2 ) \partial_l \tilde{\mu}_l  -  \tilde{\mu}_l 
 \partial_l b^2_l }{ ( b_l^2 + \tilde{\mu}_l^2 )^2 } \Bigg\},
 \label{eq:alphaflow}
 \\
 \beta_l = {} & \frac{ {u}_l}{2 ( 1 + {r}_l )}
 \Biggl\{ \frac{ b_l}{ ( b_l^2 + \tilde{\mu}_l^2 )}
 \left[  \eta_l  - \gamma_l  -  \frac{ \partial_l {r}_l }{ 1 + {r}_l } \right]
 \nonumber
 \\
 & + \frac{ ( b_l^2 - \tilde{\mu}_l^2 ) \partial_l b_l  
 +  b_l \partial_l \tilde{\mu}^2_l }{ ( b_l^2 + \tilde{\mu}_l^2 )^2 } \Bigg\}.
 \label{eq:betaflow}
 \end{align}
Finally, the bosonic anomalous dimension is
\begin{equation}
 \gamma_l 
 =  p {u}_l  \left[ \frac{ ( 1 - \eta_l ) ( b_l^2 - \tilde{\mu}_l^2)}{ ( b_l^2 + \tilde{\mu}_l^2 )^2 }
 + {\rm Re}   \frac{ \partial_l ( \tilde{\mu}_l + i b_l ) }{ ( \tilde{\mu}_l + i b_l )^3 }
 \right].
 \label{eq:gammaflow}
 \end{equation}

\section{Non-Fermi liquid fixed point}
\label{sec:fixedpoint}

To find possible fixed points of the above system of differential equations, we note that
the flow equation \eqref{eq:uflow} for the coupling $u_l$ implies that 
at
a non-trivial fixed point with
$ \lim_{ l \rightarrow \infty} u_l = u_{\ast} \neq 0$  
the fermionic and bosonic anomalous dimensions
must satisfy the scaling relation
 \begin{equation}
 \gamma_{\ast} = 2 - 2 \eta_{\ast}.
 \end{equation}
Using this to eliminate $\gamma_{\ast}$ in favor of $\eta_{\ast}$
we find that the fixed point values 
$\mu_{\ast}, {r}_{\ast} , a_{\ast}, b_{\ast}, \eta_{\ast}$, and $u_{\ast}$
of the six scale-dependent parameters $\mu_l, r_l, a_l , b_l , \eta_l$, and $u_l$ 
are constrained  by the following six equations,
 \begin{subequations}
 \label{eq:allfix}
 \begin{align}
  ( 1 - \eta_\ast ) \mu_\ast  & =  
 \frac{   u_\ast ( \mu_\ast + a_{\ast} )}{(1 + {r}_\ast  ) [  b_{\ast}^2 + (\mu_{\ast} + a_\ast )^2 ] }, \hspace{7mm}
 \label{eq:mufix}
 \\ 
 ( 2  \eta_\ast -1 ){r}_\ast 
 & =    p u_\ast \frac{ b_\ast^2 - ( \mu_\ast + a_{\ast} )^2}{ [ b_\ast^2 + ( \mu_\ast + a_{\ast})^2 ]^2 },
 \label{eq:rfix}
  \\
 \eta_{\ast} a_{\ast} & = \frac{ \eta_{\ast} u_{\ast} ( \mu_{\ast} + a_{\ast})   }{2 ( 1 + {r}_{\ast} )
 [  b_{\ast}^2 + ( \mu_{\ast} + a_{\ast} )^2] },
 \label{eq:afix}
 \\
\eta_{\ast} b_{\ast} & = \frac{  ( 3 \eta_{\ast} -2 )u_{\ast} b_{\ast}   }{2 ( 1 + {r}_{\ast} )
 [  b_{\ast}^2 + ( \mu_{\ast} + a_{\ast} )^2 ] },
  \\
 ( 1- \eta_{\ast}  )  & =   ( 1- \eta_{\ast} )    \frac{p u_{\ast}}{2} 
 \frac{ b_{\ast}^2  -( \mu_{\ast} + a_{\ast} )^2}{[ b^2_{\ast}
 +  ( \mu_{\ast} + a_{\ast} )^2 ]^2 }, 
 \hspace{10mm}
 \\
 a_{\ast}^2 + b_{\ast}^2 & = 1.
 \end{align}
 \end{subequations}
We have analytically determined the solutions of this system of equations.
For $u_{\ast} > 0$ we find physically acceptable solutions only for 
$\mu_{\ast} = a_{\ast} =0$ implying
$b_{\ast} =1$. Actually, if we allow for (unphysical) complex values of $\eta_{\ast}$
the above system has additional solutions where $\mu_{\ast}$,  $a_{\ast}$, and $b_{\ast}$ are all
finite. Moreover, for negative $u_{\ast}$ we find additional solutions 
with $b_{\ast} =0$, $a_{\ast} = \pm 1$, and
finite $\mu_{\ast}$ which we do not further discuss in this work \cite{footnoteu}.
To determine the values of $\eta_{\ast}$, $r_{\ast}$, and $u_{\ast}$ at the physical fixed point, 
we set $\mu_{\ast} = a_{\ast} =0$ and $b_{\ast} =1$ in Eqs.~\eqref{eq:allfix} and obtain
 \begin{subequations}
\begin{align}
 u_{\ast} & = \frac{2}{p},
 \label{eq:ufix}
 \\
 {r}_{\ast} & = \frac{2}{ 2 \eta_{\ast} -1},
  \\
 p \eta_{\ast} & = \frac{ 3 \eta_{\ast} - 2}{ 1 + r_{\ast}} =
\frac{ 3 \eta_{\ast} -2}{ 1 + \frac{2}{2 \eta_{\ast} -1 }}.
 \label{eq:etaquad}
 \end{align}
 \end{subequations}
The resulting 
quadratic equation for $\eta_{\ast}$ has the two solutions
\begin{subequations}
	\label{eq:etaroot}
	\begin{align}
		\eta_{\ast}^+ & =
		\frac{ 7 + p + \sqrt{ 1 + 30 p  + p^2 } }{ 4 ( 3 - p) } 
		, \\
		\eta_{\ast}^- & =
		\frac{ 4 }{ 7 + p + \sqrt{ 1 + 30 p  + p^2 } } 
		.
		\label{eq:etarootphys}
	\end{align}
\end{subequations}
%
%
%
%
%
%
\begin{figure}[tb]
 \begin{center}
  \centering
 \vspace{7mm}
  \includegraphics[width=0.45\textwidth]{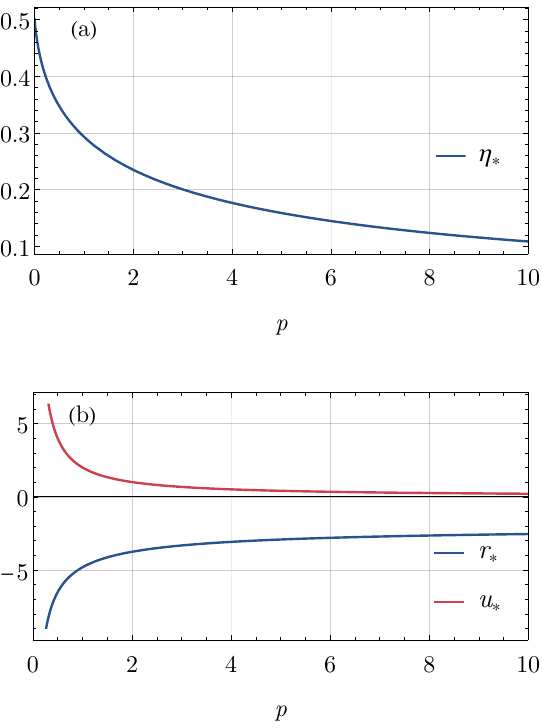}
   \end{center}
  \caption{
(a) 
Physical branch $\eta_{\ast} \equiv \eta_{\ast}^{-}$ 
of the fermionic anomalous dimension at the non-Fermi liquid fixed point
given in Eq.~\eqref{eq:etarootphys}.
(b) Corresponding fixed point values of $r_{\ast}$ (blue) and $u_{\ast}$
(red).}
\label{fig:etaplot}
\end{figure}
We discard the $\eta_{\ast}^{+}$ solution because it exhibits a singularity at $p = 3$
that we believe to be unphysical.
The only physical solution 
of our fixed point equation \eqref{eq:etaquad}
for the fermionic anomalous dimension is therefore
$\eta_{\ast} \equiv \eta_{\ast}^{-}$ given in Eq.~\eqref{eq:etarootphys},
which is shown in Fig.~\ref{fig:etaplot}(a).
Obviously, $\eta_{\ast} (p)$ is a continuous function of $p$ for all $p$ with  
 \begin{equation}
 \eta_{\ast} = \frac{1}{2} - p + {\cal{O}} ( p^2 ),
 \end{equation}
for small $p$, while for large $p$ the leading asymptotics is
 \begin{equation}
 \eta_{\ast} =\frac{2}{p} + {\cal{O}} ( 1 /p^2 ).
 \end{equation}
The corresponding fixed point values of $r_{\ast}$ and $u_{\ast}$ 
are shown in Fig.~\ref{fig:etaplot}(b).
Note that $1 + r_{\ast}$ is negative for all $p$ and diverges for $p \rightarrow 0$ as
$r_{\ast} \sim - 1/p = - u_{\ast}/2$. 
For $0 < p \ll 1$ both  $ | r_{\ast}|$ and $u_{\ast}$ 
are large compared with unity, indicating the non-perturbative nature of the non-Fermi liquid fixed point in this regime.
Given the fact that $\gamma_{\ast} = 2 - 2 \eta_{\ast} > 1$,
we conclude that for small imaginary frequencies
the boson propagator scales as
  \begin{equation}
 F ( \Omega ) \sim -  k_{\ast}  | \Omega |^{\gamma_{\ast} -1 } ,
 \label{eq:FOmega}
 \end{equation}
with some positive real constant $k_{\ast}$. 
For $ p \rightarrow \infty$ where $ \gamma_{\ast} \rightarrow 2$
this implies $ F ( \Omega ) \sim - k_{\ast} | \Omega |$, so that the analytic continuation from the
upper frequency plane to real frequencies 
($ | \Omega | \rightarrow - i \Omega $) gives for the retarded boson propagator
 \begin{equation}
 F_{\rm ret} ( \Omega + i 0) \sim i k_{\ast}   \Omega .
 \end{equation}
The resulting spectral function satisfies
  \begin{equation}
  \Omega {\rm Im} F_{\rm ret} ( \Omega + i 0) \geq 0,
 \label{eq:spectral}
 \end{equation}
which is a  general property of any bosonic spectral function. 
For finite $p$ where $1 < \gamma_{\ast} < 2$ we have to choose the physical Riemann sheet of the multi-valued complex function $z^{\gamma_{\ast} -1 }$
to obtain the physical spectral function that satisfies the positivity 
condition (\ref{eq:spectral}), see  Ref.~[\onlinecite{Schoenhammer98}] for a careful discussion.

\begin{figure}[tb]
 \begin{center}
  \centering
  \includegraphics[width=\linewidth]{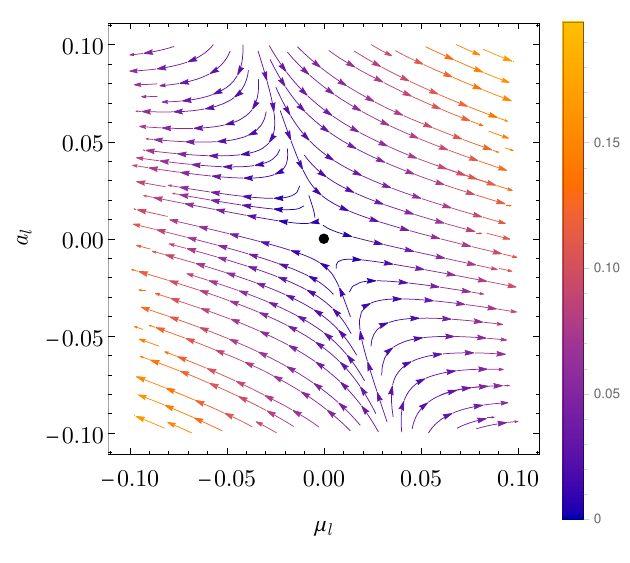}
   \end{center}
  \caption{
RG flow in the $\mu-a$ plane for $ p = 1 $ 
close to the fixed point $\mu_{\ast} = a_{\ast}=0$
obtained from the linearized flow equations \eqref{eq:linflowmua}. Note that the physical initial condition corresponding to the bare action is given by the horizontal line $a_0=0$.
}
\label{fig:muaflow}
\end{figure}

\begin{figure}[tb]
 \begin{center}
  \centering
 \vspace{7mm}
  \includegraphics[width=0.45\textwidth]{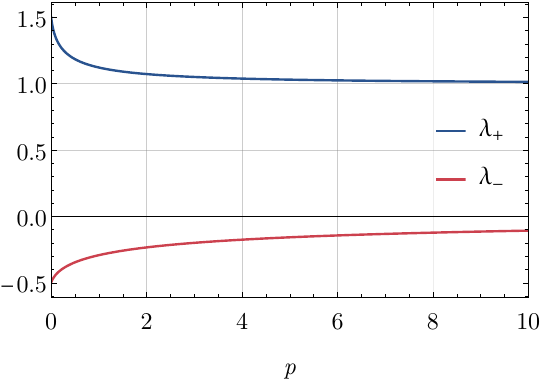}
   \end{center}
  \caption{
Graph of the eigenvalues $\lambda_{+}$ (blue) and $\lambda_-$ (red) 
of the matrix in  Eq.~\eqref{eq:linflowmua} which characterize the 
linearized RG flow in the $ \mu-a$-plane 
around the fixed point $\mu_{\ast} = a_{\ast}=0$.
}
\label{fig:eigenpmplot}
\end{figure}

To investigate the stability of the non-Fermi liquid fixed point, we now
study the linearized RG flow in the vicinity of this fixed point. 
Given the fact that at the fixed point $\mu_{\ast} = a_{\ast} =0$ we find from
Eqs.~(\ref{eq:muflow}, \ref{eq:aflow}, \ref{eq:alphaflow}) that the linearized flow in the $\mu - a$-plane decouples from the flow of the other parameters,
 \begin{equation}
 \partial_l 
\begin{pmatrix}
 \mu_l \\  a_l 
\end{pmatrix} =
 \begin{pmatrix} M_{\mu \mu} & M_{\mu a} \\
  M_{a \mu} & M_{aa} \end{pmatrix} 
  \begin{pmatrix}
  \mu_l \\ a_l \end{pmatrix} ,
 \label{eq:linflowmua}
 \end{equation}
with
 \begin{subequations}
 \begin{align}
 M_{\mu \mu } & = 1 - \eta_{\ast} - \frac{ u_{\ast} }{ 1 + r_{\ast}},
 \\
 M_{ \mu a} & = - \frac{ u_{\ast} }{ 1 + r_{\ast}},
 \\
 M_{ a \mu} & = 
 u_{\ast} \left(  
 \frac{ 1 }{ 1 + r_{\ast} } - \frac{ 1 }{ 2 ( 1 + r_{\ast} ) - u_{\ast} }
 \right) ,
 \\
 M_{ a  a } & = - \eta_{\ast} -
 \frac{ u_{\ast}^2 }{ (1 + r_{\ast} ) [2 ( 1 + r_{\ast} ) - u_{\ast}] } .
 \end{align} 
 \end{subequations}
The resulting RG flow in the $ \mu-a$-plane is shown in
Fig.~\ref{fig:muaflow}.
Obviously, the RG flow in the $\mu-a$-plane has one attractive and one repulsive direction.
To characterize the behavior of the RG trajectories quantitatively, we calculate the eigenvalues
$\lambda_{\pm}$ of the $2 \times 2$ matrix
in Eq.~\eqref{eq:linflowmua},
 \begin{equation}
 \lambda_{ \pm} =  \frac{ M_{\mu \mu} + M_{ a a}}{2}
 \pm \sqrt{ \left( \frac{ M_{ \mu \mu} - M_{aa}}{2} \right)^2 + M_{ \mu a} M_{ a \mu } }.
 \end{equation}
At small and large p,
they behave as
\begin{subequations}
	\begin{align}
		\lambda_+ & = \frac{3}{2} - 3 p + {\cal O}(p^2) , \\
		\lambda_- & = - \frac{1}{2} + p + {\cal O}(p^2) ,
	\end{align}
\end{subequations}
and
\begin{subequations}
	\begin{align}
		\lambda_+ & = 1 + {\cal O}(1/p^2) , \\
		\lambda_- & = - \frac{2}{p} + {\cal O}(1/p^2) ,
	\end{align}
\end{subequations}
respectively.
We plot these eigenvalues as function of $p$ in Fig.~\ref{fig:eigenpmplot}.

\begin{figure}[tb]
 \begin{center}
  \centering
 \vspace{7mm}
  \includegraphics[width=0.45\textwidth]{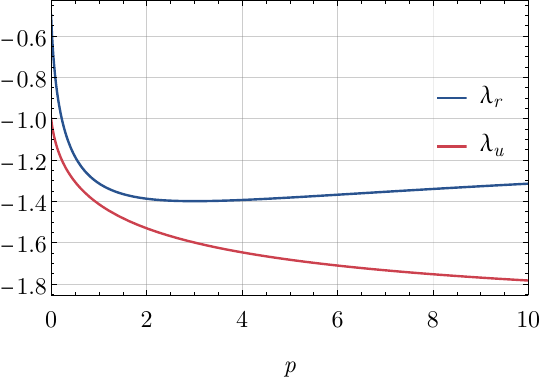}
   \end{center}
  \caption{
Graph of the eigenvalues $\lambda_r$ (blue) 
and $\lambda_u$ (red) 
of the linearized flow in the vicinity of the non-Fermi liquid fixed point, see
Eqs.~\eqref{eq:lambdau} and~\eqref{eq:lambdardef}.
}
\label{fig:lambdaurplot}
\end{figure}

\begin{figure}[tb]
 \begin{center}
  \centering
  \includegraphics[width=\linewidth]{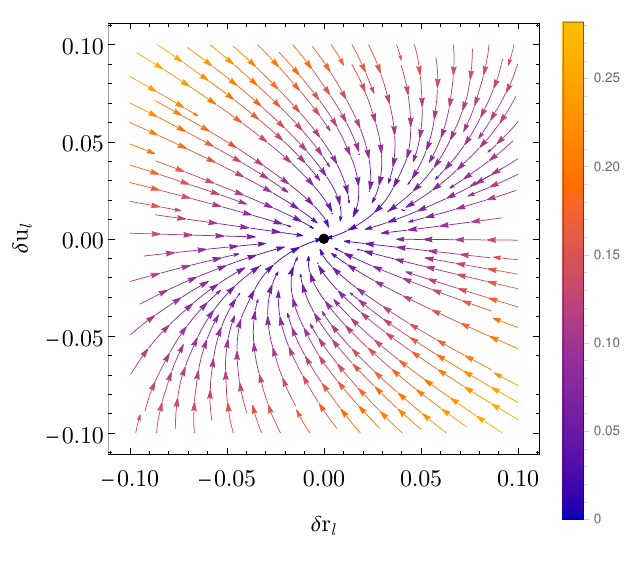}
   \end{center}
  \caption{
RG flow in the $r-u$-plane for $ p = 1 $
close to the non-Fermi liquid fixed point
obtained from the linearized flow equations 
\eqref{eq:ulinflow} and \eqref{eq:rlinflow}.
}
\label{fig:urflow}
\end{figure}

Next, let us discuss the linearized RG flow of the couplings $u_l$ and $r_l$ in the
vicinity of our non-Fermi liquid fixed point.
Using  Eqs.~(\ref{eq:uflow}, \ref{eq:betaflow}, \ref{eq:gammaflow})
we find that the linearized flow of $ \delta u_l = u_l - u_{\ast}$ completely 
decouples from the other
parameters,
 \begin{equation}
 \partial_l \delta u_l = - p u_{\ast} 
  ( 1 - \eta_{\ast} ) \delta u_l = \lambda_u \delta u_l,
 \label{eq:ulinflow}
 \end{equation}
with 
\begin{equation}
 \lambda_u = - p u_{\ast}  ( 1 - \eta_{\ast}) 
= -2 ( 1 - \eta_{\ast} ),
 \label{eq:lambdau}
 \end{equation}
where we have used  Eq.~\eqref{eq:ufix} to set
$p u_{\ast} = 2$.
Given the fact that $ 0 < \eta_{\ast} < 1/2$, 
we conclude that the coupling $u_l$ is 
irrelevant at the non-Fermi liquid fixed point 
with scaling exponent $\lambda_u < 0$.
A graph of $\lambda_u$ as a function of $p$
is shown by the red line in Fig.~\ref{fig:lambdaurplot}.
Finally, using Eqs.~(\ref{eq:rflow}, \ref{eq:betaflow}, \ref{eq:gammaflow})
we obtain for the linearized flow of $\delta r_l = r_l - r_{\ast}$,
 \begin{equation}
 \partial_l \delta r_l = \lambda_r \delta r_l + \lambda_{ ru} \delta u_l ,
 \label{eq:rlinflow}
 \end{equation}
where
 \begin{subequations}
 \begin{align}
 \lambda_r & = 
 2 \eta_{\ast} - 1 - 
 \frac{ 
 p r_{\ast} u_{\ast}^2 \left( 5 \eta_{\ast} - 3 \right)
 }{  
 2 ( 1 + r_{\ast} )^2 - ( 1 + r_{\ast} ) u_{\ast} - p u_{\ast}^2 }
 ,
 \label{eq:lambdardef}
 \\
 \lambda_{ r u} & = 
 - p 
 \nonumber\\
 & \hspace*{-5mm}
 + p r_{\ast} \frac{
 	p u_{\ast}^2 + 2 ( 1 + r_{\ast} )^2 ( \eta_{\ast} - 1 )
 	+ u_{\ast} ( 1 + r_{\ast} ) ( 2 \eta_{\ast} - 1 )
 }{
 2 ( 1 + r_{\ast} )^2 - ( 1 + r_{\ast} ) u_{\ast} - p u_{\ast}^2 }
 .
 \end{align}
 \end{subequations}
The eigenvalue $\lambda_r$ as a function of $p$ is represented by the
blue line in Fig.~\ref{fig:lambdaurplot}.
It is always negative and has the asymptotics
\begin{subequations}
	\begin{align}
		\lambda_r & = - \frac{1}{2} - 6 p + {\cal O}( p^2 ) , \\
		\lambda_r & = - 1 - \frac{ 8 }{ p } + {\cal O}( 1/p^2 )
	\end{align}
\end{subequations}
for small and large $p$, respectively.
A projection of the linearized RG flow onto the plane spanned by
$\delta r_l$ and $\delta u_l$
in the vicinity of the fixed point is shown in Fig.~\ref{fig:urflow}.
Note that the projected flow has only attractive directions,
so that  fine tuning of the couplings
 $r_l$ and $u_l$ is not necessary to
realize the critical state associated with the non-Fermi liquid fixed point. 
Our model therefore exhibits self-tuned criticality \cite{Pan21}
with respect to the bosonic mass parameter $\Delta + \Pi ( 0)$.  
Thus, the only relevant coupling at the
non-Fermi liquid fixed point is the scaling variable associated with the positive
eigenvalue $\lambda_+$ of the $ 2 \times 2$-matrix in Eq.~\eqref{eq:linflowmua}, which is a linear combination of the
rescaled chemical potential $\mu_l$ and the spectral asymmetry parameter $a_l$.

\section{Summary and conclusions}

\label{sec:conclusions}

In this work we have developed a FRG approach for a
dissipative  Yukawa-SYK model 
where the inverse boson propagator 
exhibits a non-analytic $ | \Omega |$ frequency dependence. 
We have shown that, to leading order in $1/N$ and $1/M$,
the infinite hierarchy of
FRG flow equations for this model
can be reduced to a system of flow equations
for the irreducible fermionic and bosonic self-energies and two types of scale-dependent 
four-point vertices. 
This system is closed because the flow of the four-point vertices can be expressed again 
in terms of the self-energies. 
Within a standard low-energy expansion of the self-energies
 we have found a non-trivial non-Fermi liquid fixed point with  
critical exponents depending on the ratio $N /M$.
A stability analysis of the linearized RG flow in the vicinity of this fixed point shows that 
it has  only one repulsive direction corresponding to a linear combination
of the rescaled chemical potential $\mu_l$ and a parameter $a_l$ which quantifies the 
spectral asymmetry. As $a_0=0$ in the microscopic action, the physical parameter that can be tuned to reach the non-Fermi liquid fixed point is the fermionic density.  
In particular, the rescaled boson mass parameter $r_l$ and the Yukawa coupling $u_l$ are
both irrelevant at the fixed point, so that no fine-tuning of these parameters is necessary
to realize the corresponding non-Fermi liquid phase. 
Although in principle it should also be possible to extract these results from the
corresponding Dyson-Schwinger equations, in practice our approach based on FRG flow equations is more convenient because it allows us to extract the low-energy properties analytically using
well-established approximations.

It would be interesting to extend our analysis to the usual YSYK model where the inverse boson propagator exhibits a quadratic frequency dependence. Our FRG flow equations (\ref{eq:flowselfaverage2}--\ref{eq:flow4b})
remain valid also in  this case, so that from a numerical solution of these equations we expect to recover the
non-Fermi liquid solution of the Dyson-Schwinger equations derived in Ref.~[\onlinecite{Classen21}]. Unfortunately, the low-energy
expansion of Sec.~\ref{sec:flow}, which is crucial to make progress analytically, does not produce
sensible results in this case, at least when it is combined with a sharp frequency regulator. Possibly,
an ultra-smooth regulator of the type proposed by Husemann and Salmhofer \cite{Husemann09} might solve this problem.

	\section*{ACKNOWLEDGEMENTS}
We thank
 Olexandr Tsyplyatyev for useful discussions and the 
Deutsche Forschungsgemeinschaft (DFG, German Research Foundation) for financial support via  TRR 288 - 422213477.

%


%

\end{document}